\documentclass[a4paper,11pt]{article}
\pdfoutput=1 
\usepackage{jcappub}
\usepackage{graphicx, bbm, slashbox}
\usepackage{epsfig}
\usepackage{bm}

\newcommand\T{\rule{0pt}{2.6ex}}       
\newcommand\B{\rule[-1.2ex]{0pt}{0pt}} 

\title{High-Energy Neutrino Signatures of Newborn Pulsars In the Local Universe}

\author{Ke Fang}
\affiliation{Department of Astronomy \& Astrophysics, Kavli Institute for Cosmological Physics, The
  University of Chicago, Chicago, Illinois 60637, USA.}

\abstract{
Charged particles can be accelerated  to higher than PeV energies  in the electro-magnetic wind of a  fast-spinning newborn pulsar to produce high-energy neutrinos, through hadronuclear interactions in the supernova remnant.  Here we explore the detectability and observational signatures of these high-energy neutrinos. We show that their spectral index  varies approximately from 1.5 to 2, depending on the relevant pulsar properties and observation time. We also apply the scenario to existing young pulsars in the local universe and find the corresponding neutrino flux well below current detection limits. Finally, we estimate the birth rate of  fast-spinning pulsars in the local Universe that can be observed by the IceCube observatory to be  0.07 per year, with an upper limit of 0.29 per year. 
}

\begin{document}
\maketitle

\section{\label{sec:intro} Introduction}

Fast-spinning pulsars are known to release their rotational energy via electromagnetic radiation \citep{1969ApJ...157.1395O}. Charged particles, including heavy ions \citep{Hoshino92, Gallant94}, can be accelerated by energy conversion of the wind Poynting flux into kinetic energy in pulsar winds \citep{PhysRevLett.22.728, Blasi00, Arons03}. When cosmic rays escape from the acceleration site, they interact with the surrounding supernova ejecta \citep{FKO12}.  High-energy neutrinos are then produced during hadronuclear interactions.  Such high-energy neutrinos are among the first multimessenger signals released by the star, and  in addition, they experience no delay in galactic and intergalactic magnetic fields. Hence a measurement of the high-energy neutrino signal from a future event can provide a unique way to identify the birth of a newborn pulsar and the start of high-energy acceleration in that system.

The IceCube Observatory recently reported the detection of a diffusive flux of high-energy  astrophysical neutrinos \cite{Aartsen:2013jdh, Aartsen:2014gkd}. These 37  events have deposited energies between $\sim 50\,\rm TeV$ and $2\,\rm PeV$,  and are consistent with an isotropic $E^{-2}$ spectrum. Meanwhile, searches for point-like neutrino sources, applying four years of IceCube data, still find the event data to be compatible with  the background signal \cite{0004-637X-779-2-132,Aartsen:2014cva}.  No point sources of high-energy neutrinos have been identified thus far. 

We showed that a diffusive EeV ($10^{18}\,\rm eV$) neutrino flux should be effectively produced if extragalactic newborn pulsars are the emitters of ultrahigh energy cosmic rays \citep{Fang:2013vla}.  
In this work we consider, more generally, the TeV ($10^{12}\,\rm eV$) to EeV 
high-energy neutrino emissions  from  individual fast-spinning pulsars in the local universe, focusing on their light curve, energy spectrum, and flux detectability, especially in light of  recent IceCube measurements. In particular, we extend the previous work of \cite{Murase09}  to a study with complete pulsar population including magnetars and Crab-like pulsars, and update those results based on a full Monte Carlo simulation. We also apply the scenario to existing nearby pulsars and  find  their emission levels to be consistent with the non-detection. Based on the discovery sensitivities of IceCube, we provide a prediction of the detectability and a template of emission profiles for high-energy neutrinos from a future nearby newborn pulsar. 

Section~\ref{Sec:theory} describes  the neutrino production mechanism. Following a description of the numerical setup of this work in Section~\ref{Sec:code}, the results of time-integrated neutrino spectrum is presented in Section~\ref{Sec:timeBin}, focusing on the effects from secondary interactions and the dependence on pulsar properties. Section~\ref{Sec:flux} presents the time evolution  of the neutrino emissions, and  provides a template for future  detections to identify local pulsar sources. We also apply the scenario to recent nearby pulsar systems and test the agreement to the non-detections of point sources. In Section~\ref{Sec:flux_rate} we investigate the birth rate of future pulsars that could be detected by the IceCube Observatory.   We draw conclusions and discuss the implications for future neutrino observations in Section~\ref{Sec:discussion}.

\section{Neutrino production mechanism} \label{Sec:theory}
The idea that cosmic rays could be accelerated by rotating neutron stars dates back to \cite{1969ApJ...157.1395O}. \cite{Blasi00, Arons03} further developed the scheme of the production of ultra-high energy cosmic ray in the electric field associated to the rotating magnetic dipole,  although uncertainties still exist especially regarding the seeding of heavy ions \citep{Hoshino92, Gallant94} and the acceleration sites \citep{Arons03, FKO12}. Here we review the cosmic ray production process with energy dissipation by electron pairs additionally taken into account, and calculate the corresponding neutrino emissions. 

Considering a newborn pulsar spinning down and releasing rotational energy by electromagnetic radiation, its characteristic spin-down time reads $\tau_{\rm EM} = (1/2\,I\,\Omega_i^2)/\dot{E}_{\rm rot} = 1\,B_{13}^{-2}\,P_{i,-3}^2\,\rm yr$, where  $\dot{E}_{\rm rot} = 4\,\mu^2\,\Omega_i^4/9\,c^3$ is the dipole radiation rate  for a pulsar with inertia $I\approx 10^{45}\,\rm g\,cm^2$,  surface magnetic field $B=10^{13}\,B_{13}\,\rm G$ and initial spin period $P_i = 1\,P_{i,-3}\,\rm ms$. Please note that in the following text, two corresponding parameters - magnetic moment $\mu=BR^3/2 \approx 10^{31}\,B_{13}\,\rm erg/G$ and initial rotation speed $\Omega_i=2\pi/P_i = 10^{3.8}\,P_{i,-3}^{-1}\,\rm s^{-1}$, might be used as  alternatives to  $B$ and $P$.
We assume that the electromagnetic energy is efficiently converted into the kinetic energy of charged particles including ions and electrons, with a rest frame injection rate  \citep{LKP13}
\begin{equation}
\dot{N}mc^2 =\frac{ \dot{N}_{\rm GJ}}{Z}\,m_i c^2 + 2\kappa\dot{N}_{\rm GJ}m_e c^2
\end{equation}
Here $\dot{N}_{\rm GJ}= \sqrt{\dot{E}_{\rm rot}\,c}/e = \mu\,\Omega^2/ec$ is   the Goldreich-Julian  rate  \cite{Goldreich69}, that is the number of elementary charges per unit volume that enter the parallel electric fields in the magnetosphere, $m_i=Am_p$ and $Z$ are the mass and the charge of the ion, and $\kappa$ is the pair-production multiplicity (see for e.g. \cite{2001ApJ...560..871H}). Let $y_i \equiv 2Z\kappa m_e/m_i$, which can be interpreted as the ratio of the kinetic energy that goes into electrons to that goes into ions.  In crab-like pulsars $\kappa\sim10^4$ \citep{2010arXiv1005.1831B}, then $y_i\sim 10$.

If a fraction $\eta$ of the electromagnetic luminosity   turns into the kinetic luminosity, an ion injected at system time $t$ would obtain  energy of 
\begin{equation}\label{eqn:Et}
E_{\rm CR} (t) = \frac{\eta\, \dot{E}_{\rm rot}(t) } {\dot{N}mc^2} \, m_i c^2 =   E_{\rm CR, max}\,\left(1+{t}/{\tau_{\rm EM}}\right)^{-1} 
\end{equation}
where
\begin{eqnarray}\label{eqn:Emax}
E_{\rm CR, max}&=&
\eta Ze\,\frac{\Omega_i^2\mu}{c^2}\,\frac{1}{1 + y_i} \\ \nonumber
&\approx & 1.7\times 10^{18}\,B_{13}\,P_{i,-3}^{-2}\,\eta_{-0.5}\,A\,\kappa_4^{-1} \,\rm eV\quad\quad y_i\gg1
\end{eqnarray} 
is the maximum energy  the ion can gain from the induced electric potential \citep{Blasi00, Arons03}. Note that we have taken an acceleration efficiency $\eta=0.3\,\eta_{-0.5}$ as suggested by  \cite{FKO13,2010arXiv1005.1831B}.

Notice that the conduction currents in the pulsar wind have a charge density proportional to the particle energy $\dot{N}_{\rm GJ}\propto E_{\rm CR}$. As a consequence,  the  prompt particle injection rate  is  a constant over time:
\begin{equation}\label{eqn:dNdEdt}
\frac{dN_{\rm CR}}{dE\,dt}=\frac{c}{(Ze)^2\,\eta}\,(1+y_i)
\end{equation}
Integrating it over  the corresponding acceleration period $\Delta t (E) =  (E_{\rm rot} \eta/(1+y_i)) / \dot{E}_{\rm rot}$, one gets  the time-integrated energy spectrum 
\begin{equation}\label{eqn:dNdE}
\frac{dN_{\rm CR}}{dE}=\frac{9}{8}\,\frac{c^2\,I}{Z\,e\,\mu}\,\frac{1}{E_{\rm CR}}
\end{equation}
Notice that in the above calculations we ignored the gravitational wave losses, which would only become significant for neutron stars with surface magnetic field $B\gg10^{15}\,\rm G$ \citep{Arons03, 2011PhRvD..84b3002K}. 
 
After accelerated in  the pulsar winds, particles would cross the  supernova ejecta surrounding the star at $R_{\rm ej} (t) =3\times 10^{16}\,\beta_{\rm ej, -1.5}\,t_{\rm yr}\,\rm cm$ to escape from the source. The ejecta can be modeled as a shell spherically expanding with speed  $\beta=((1/2I\Omega_i^2 + E_{\rm ej})/M_{\rm ej}\,c^2)^{1/2} = 0.03\,\beta_{-1.5}$
and column density  $\Sigma_{\rm SN}(t) = 5.3\,M_{\rm ej,1}\,\beta_{-1.5}^{-2}\, t_{\rm yr}^{-2}~{\rm g\,cm}^{-2}$ \citep{FKO12}, where $M_{\rm ej, 1}=10\,M_{\rm ej}/M_\odot$ is the ejecta mass and $E_{\rm ej}$ is the ejecta energy that includes both star's rotational and explosion energy. 

In an early environment, cosmic rays would interact with the dense ejecta through hadronuclear interactions,  $N+p\rightarrow N'+ \pi +\,\rm others$. The effective optical depth reads $\tau = n_b\,\sigma\,\kappa\,R_{\rm ej}$.
For protons, $\tau_{\rm pp} =0.16\,M_{\rm ej, 1}\,\beta_{-1.5}^{-2}\,t_{\rm yr}^{-2}$, with   $\sigma_{\rm pp}\sim10^{-25}\,\rm cm^2$,   $\kappa\sim0.5$ being the cross section and elasticity of proton-proton interaction at  EeV. For nuclei, since $\sigma_{Np} \propto A^{2/3}\,\sigma_{\rm pp}$ and $\kappa\propto 0.7/A$, the effective optical depth for nuclei ($\tau_{\rm Np}$) is reduced by $\sim A^{1/3}$. The system becomes optically thin  after $\tau_{\rm thin}\equiv t(\tau_{\rm pp}=1)=0.4\, M_{\rm ej,1}^{1/2}\,\beta_{-1.5}^{-1}\,\rm yr$.  The pion production efficiency is therefore 
\begin{equation}
f_\pi=\min(\tau_{\rm Np}, 1)
\end{equation} 

Charged pions produced in the  hardonuclear interactions have a relatively short life time $\tau_\pi=2.6\times10^{-8}\,\rm s$ in rest frame. However, at early stages the ejecta would so dense that  the pions  interact with the baryons before they decay. The $\pi p$ interaction time reads $t_{\pi p}=(n_b\,\sigma_{\pi p}\,\kappa_{\pi p}\,c)^{-1}$, with cross section $\sigma_{\pi p} = 5 \times 10^{-26}\,\rm cm^2$ and  elasticity $\kappa_{\pi p}\sim0.5$  \citep{PhysRevD.68.083001}. The probability that a $\pi p$ interaction happens before the pion decays  introduces a suppression factor, 
\begin{equation}\label{eqn:fsup}
f_{\rm sup} = 1 - e^{-t_{\pi p}/\gamma_\pi\,\tau_\pi}
\end{equation}
which can be approximated by $f_{\rm sup}\approx t_{\pi p}/\gamma_\pi\,\tau_\pi$ when  $t_{\pi p}\ll \gamma_\pi\,\tau_\pi$. Notice that the hadronic cooling of muons is negligible compared to that of pions since $\sigma_{\mu p}\approx 2\times10^{-28}\ll \sigma_{\pi p}$\citep{PhysRevD.68.083001}. Finally charged pions would decay into neutrinos via $\pi^\pm\rightarrow e^\pm+\nu_e(\bar{\nu}_e)+\bar{\nu}_\mu +\nu_\mu$ after ejecta gets thin.
After taking into account that the pion carries about $20\%$ of the proton's energy in a  pp interaction, and that each of the neutrino inherits a quarter of the charged pion’s energy, a characteristic energy of the neutrino products is  $E_\nu \approx 0.05\,E_p$. Taking along the suppression factor, the neutrino flux produced by hadronuclear interactions in the ejecta can be calculated as 
\begin{equation}
E_\nu^2 \,\Phi_\nu=\frac{3}{8}\,E_{\rm CR}^2\,\Phi_{\rm CR}\,f_\pi\,f_{\rm sup}.
\end{equation}

The  flux should reach a   peak before ejecta getting thin at $\tau_{\rm thin}$, but after the suppression from $\pi p$ interaction turning relatively weak when $t_{\pi p} = \gamma_\pi \,\tau_\pi$. Assuming $E_{\pi,\rm max}=0.2 \,E_{\rm CR, max}$, the latter condition is equivalent to 
\begin{equation}\label{eqn:tpeak}
t_{\rm peak}^3\,\left(1+\frac{t_{\rm peak}}{\tau_{\rm EM}}\right) = \frac{E_{\pi,\rm max}}{m_\pi\,c^2}\,\tau_\pi\,\frac{M_{\rm ej}\,\sigma_{\pi p}\,c}{\frac{4}{3}\pi m_b\,(\beta c)^3}
\end{equation}

Below we estimate the peak time and the peak energy with two representative cases: I) a fast-spinning magnetar with  $P_i=0.6\,P_{-3.2}\,\rm ms$ and $B = 10^{15}\,B_{15}\,\rm G$, located at a distance $D=5\,\rm Mpc$. We will call this case {\it magnetar}; II) a crab-like pulsar  with  $ B=10^{13}\,B_{13}\,\rm G$ and $P_i=20\,P_{-1.7}\,\rm ms$, located at $D=10\,\rm kpc$. We denote this case as {\it crab}. In the magnetar case, $t_{\rm peak}\gg \tau_{\rm EM}=3\times10^3\,\rm s$, equation~\ref{eqn:tpeak} raises
\begin{eqnarray}
t_{\rm peak} &=& 2.4\times10^5\, B_{15}^{-1/4}\,A^{1/4}\,\kappa_4^{-1/4}\,\eta_{-0.5}^{1/4}\,M_{\rm ej,1}^{1/4}\,P_{i,-3.2}^{3/4}\,\rm s \\
E_{\nu,\rm peak} &=& 1.1\times10^{17}\,B_{15}^{-3/4}\,\eta_{-0.5}^{3/4}\,A^{3/4}\,\kappa_4^{-3/4}\,M_{\rm ej, 1}^{-1/4}\,P_{i,-3.2}^{-3/4}\,\rm eV
\end{eqnarray} 
In contrast, in the crab case $t_{\rm peak}\ll \tau_{\rm EM}=1.2\times10^{10}\,\rm s$ instead, and
\begin{eqnarray}
t_{\rm peak} &=& 2.1\times 10^5\,B_{13}^{1/3}\,P_{i,-1.7}^{-2/3}\,A^{1/3}\,\kappa_4^{-1/3}\,\eta_{-0.5}^{1/3}\,M_{\rm ej,1}^{1/3}\,\rm s \\
E_{\nu,\rm peak} &=& 2.1\times10^{14}\,B_{13}\,P_{i,-1.7}^{-2}\,\eta_{-0.5}\,A\,\kappa_4^{-1}\,\rm eV
\end{eqnarray}

\section{Numerical Setup}\label{Sec:code}
The interactions between high-energy nuclei and ejecta baryons were simulated by Monte Carlo as we did in \cite{FKO12}. In particular, the Np and pp interactions were calculated based on the hadronic interaction model EPOS \citep{PhysRevC.74.044902} and the fragmentation model as implemented in the air shower simulation code CONEX \citep{Bergmann2007420}. For this work, in order to take into account the suppression in neutrino production, we further adapted the code with pion-proton interactions calculated using EPOS. 

\begin{figure}[h]
\centering
\epsfig{file=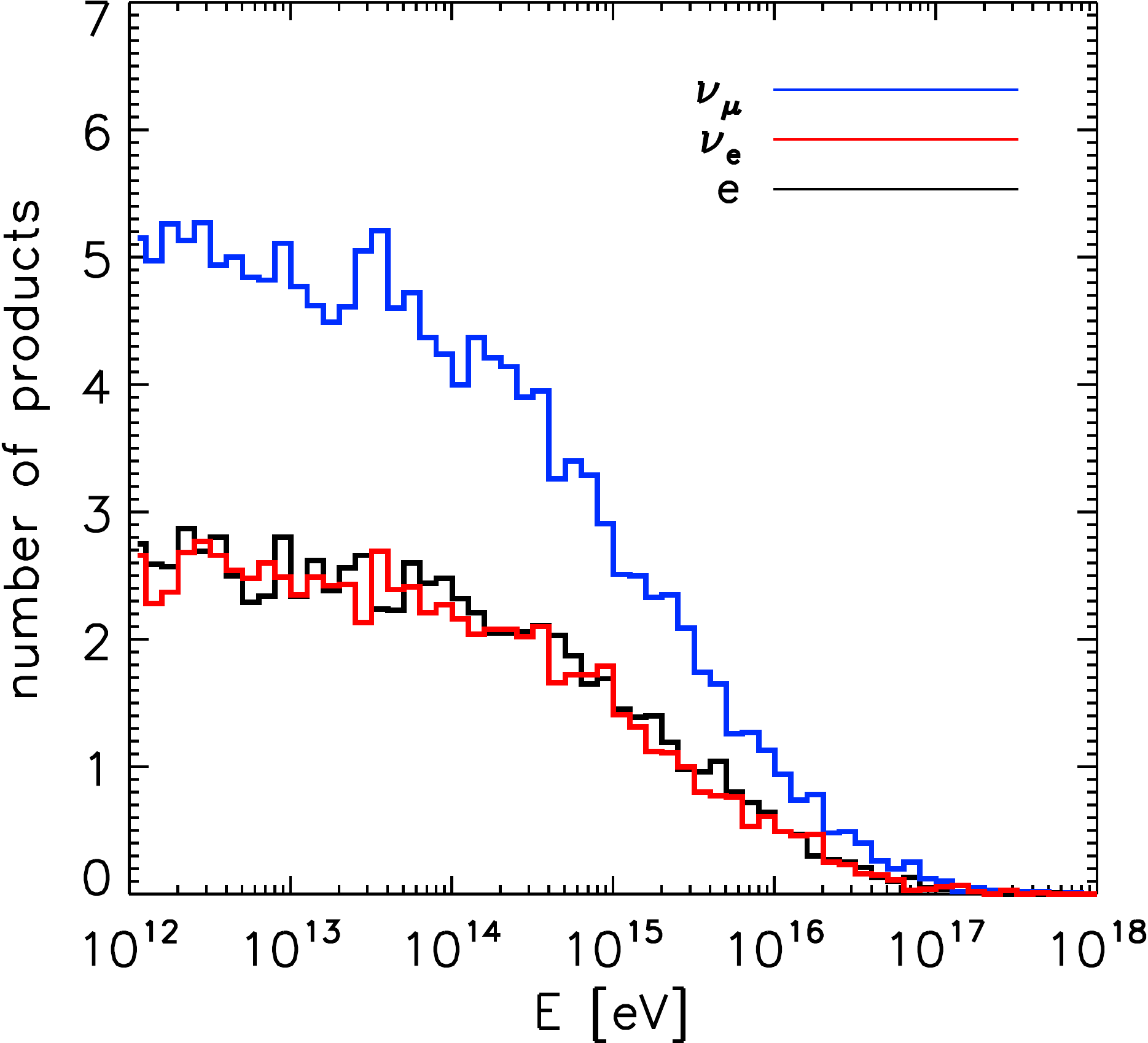,width=0.6\textwidth,clip=}
\caption{\label{fig:epos} Decay products of $\nu_e$, $\nu_\mu$ and $e$ from a $\pi p$ interaction, allowing all pions to decay without further interactions. The injected pion had energy of 1 EeV and the proton was assumed to be at rest in lab frame. The histogram is an average of 100 realizations using the hadronic interaction model EPOS  \citep{PhysRevC.74.044902}.
}
\end{figure}

In Figure~\ref{fig:epos} we show a histogram of neutrino and electron products of $\pi p$ interaction between a  1 EeV $\pi^+$ and a rest proton in the lab frame. The flux ratio between electron, electron neutrino and muon neutrino before mixing is consistent with $\Phi_e:\Phi_{\nu_e}:\Phi_{\nu_\mu}=1:1:2$ as expected from the decay of charged pions. Note that this result is obtained by allowing all intermediate pions and muons to fully decay to leptons. This might not be true in an intense interaction environment. As Equation~\ref{eqn:fsup} shows, a suppression would happen as pions are tend to interact rather than decay when the ejecta is still dense.  Remember that the elasticity of $\pi p$ interaction at EeVs is about 0.5  \citep{Murase09}, so roughly half of the pion's energy is passed to additional interactions that produce extra lower energy neutrinos.

In the simulations, we modeled pulsars  with selected initial rotation speed $\Omega_i$ and magnetic moment $\mu$.
To calculate time-integrated neutrino spectrum, we injected cosmic rays following a power-law spectrum as in Equation~\ref{eqn:dNdE}, with minimum energy $E_{\rm min} = 10^{12}\,\rm eV$ and maximum energy calculated by Equation~\ref{eqn:Emax}. To calculate time-dependent neutrino emission at chosen time $t$, we injected cosmic rays with single energy decided by Equation~\ref{eqn:Et} and flux normalized with Equation~\ref{eqn:dNdEdt}. The injected cosmic rays were then propagated through a supernova envelope of total ejected mass $10\,M_\odot$ expanding at $v_{\rm ej} = (2\,E_{\rm ej}/M_{\rm ej})^{1/2}$. The ejecta energy includes the pulsar's rotational energy and the supernova explosion energy, $E_{\rm ej} = I\Omega_i^2/2 + E_{\rm exp}$. Each time when a pion was produced, the suppression factor in Equation~\ref{eqn:fsup} was calculated to decide if it would decay or interact, by  comparing with a random number drawn from a uniform distribution between 0 and 1. All the products from primary and higher order interactions, which we will  call secondary particles in general,  were tracked down to $E_{\rm min}$. As we will see in Section~\ref{Sec:timeBin}, these secondary particles play a key role in neutrino production. The input parameters used for  the  magnetar  and  crab cases are summarized in Table~\ref{table:input}, and that for the existing nearby pulsars are listed in Table~\ref{table:pulsar}.

\begin{table}[t]
\caption{Summary of input parameters} \label{table:input}
\centering
\begin{tabular}{ccccccccc}
\hline\hline 
Model \T & $P_i$ & B & $M_{\rm ej}$ & $E_{\rm exp}$ & $\eta$ & $\kappa$ & I & Distance \\ 
& (ms)  &  (Gauss) & ($M_{\odot}$)&   ($\rm erg$) & & &($\rm g \,cm^{-2}$) & (kpc)\B  \\
\hline
Magnetar \T& 0.6 & $10^{15}$ & 10 & $10^{51}$  & 0.3 & $10^4$ & $10^{45}$ & 5000 \\

Crab & 20 & $10^{13}$ & 10  & $10^{51}$ & 0.3 & $10^4$ & $10^{45}$ & 10 \B\\ 
\hline
\end{tabular}
\end{table}

\section{Time-integrated Neutrino Spectrum}\label{Sec:timeBin}

\begin{figure}
\centering
\epsfig{file=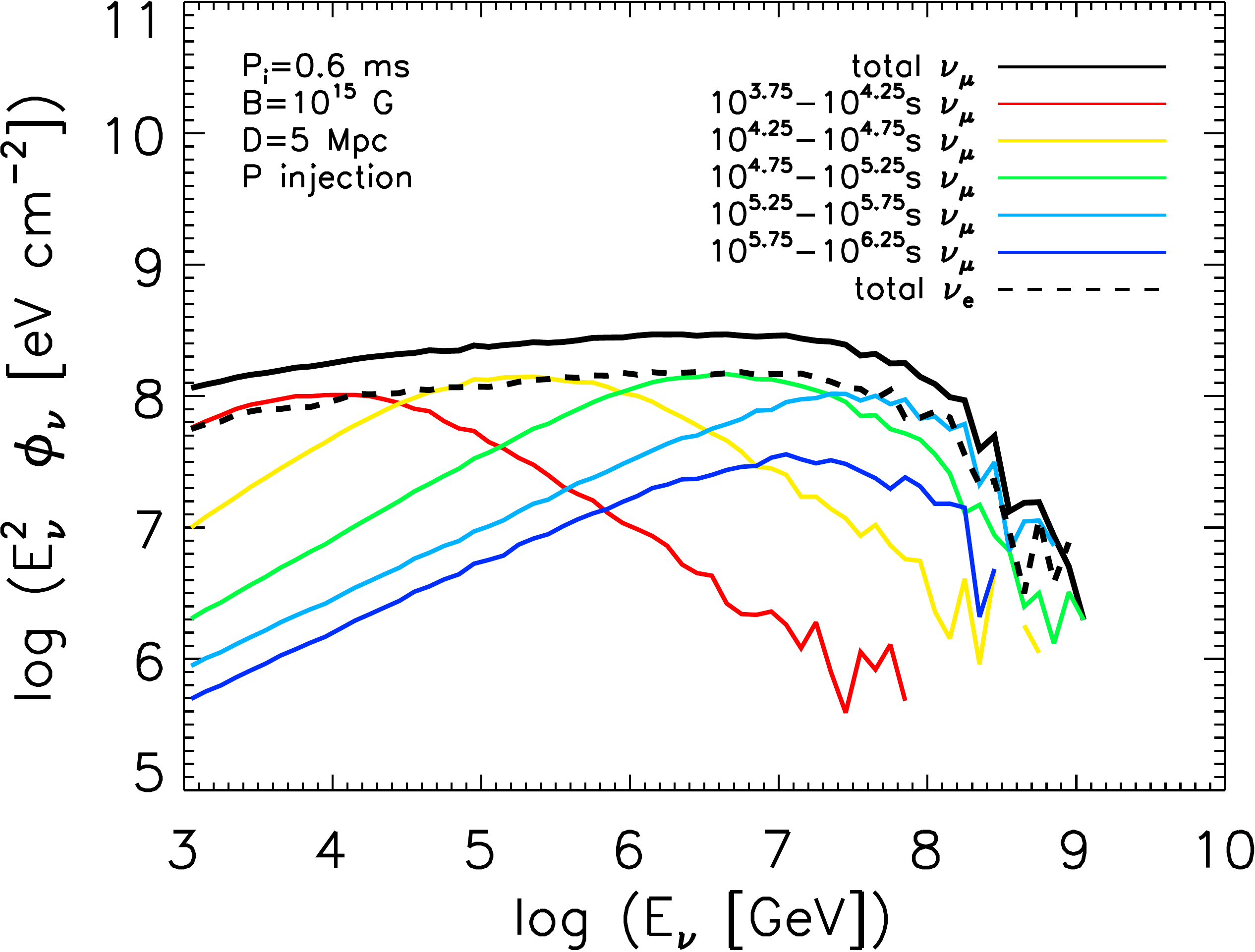,width=0.51\textwidth,clip=}
\epsfig{file=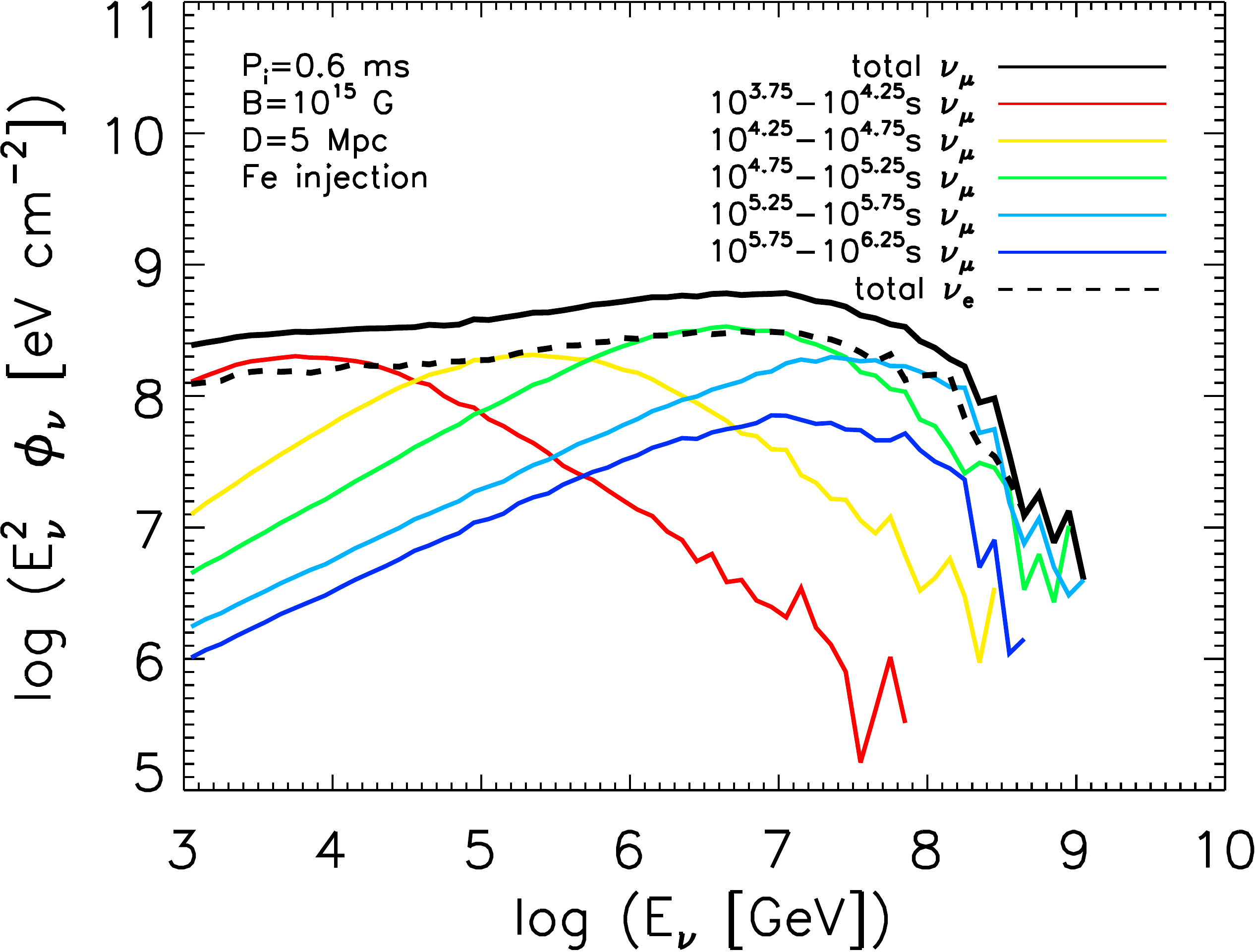,width=0.51\textwidth,clip=}
\epsfig{file=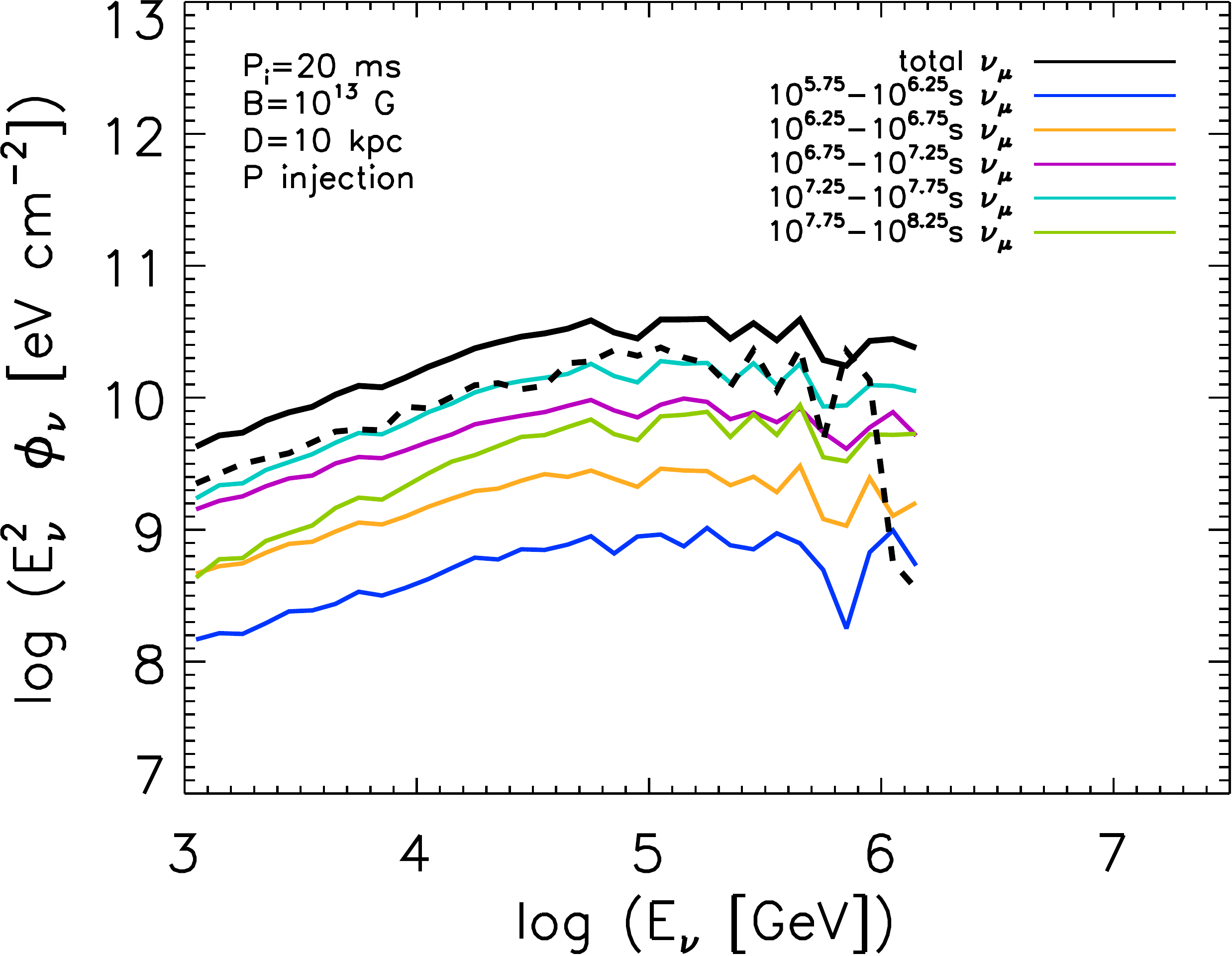,width=0.51\textwidth,clip=}
\caption{\label{fig:spec} Time-integrated spectrum of muon neutrinos (solid) and electron neutrinos (dash) from newborn pulsars in the local universe. Top and middle: magnetar case with injected cosmic rays being pure proton (top) and iron nuclei (middle) - the source is a fast-spinning magnetar with  $P_i=0.6\,\rm ms$ and $B = 10^{15}\,\rm G$ located at $D=5\,\rm Mpc$.  Bottom: crab case with proton injection - the source is a crab-like pulsar with $P_i=20\,\rm ms$ and  $ B=10^{13}\,\rm G$  located  at $D=10\,\rm kpc$. The distances of the sources were artificially chosen to fit the plotting scale. The neutrino contributions from different time intervals are listed as in the legend box.}
\end{figure}

\begin{figure}[h]
\centering
\epsfig{file=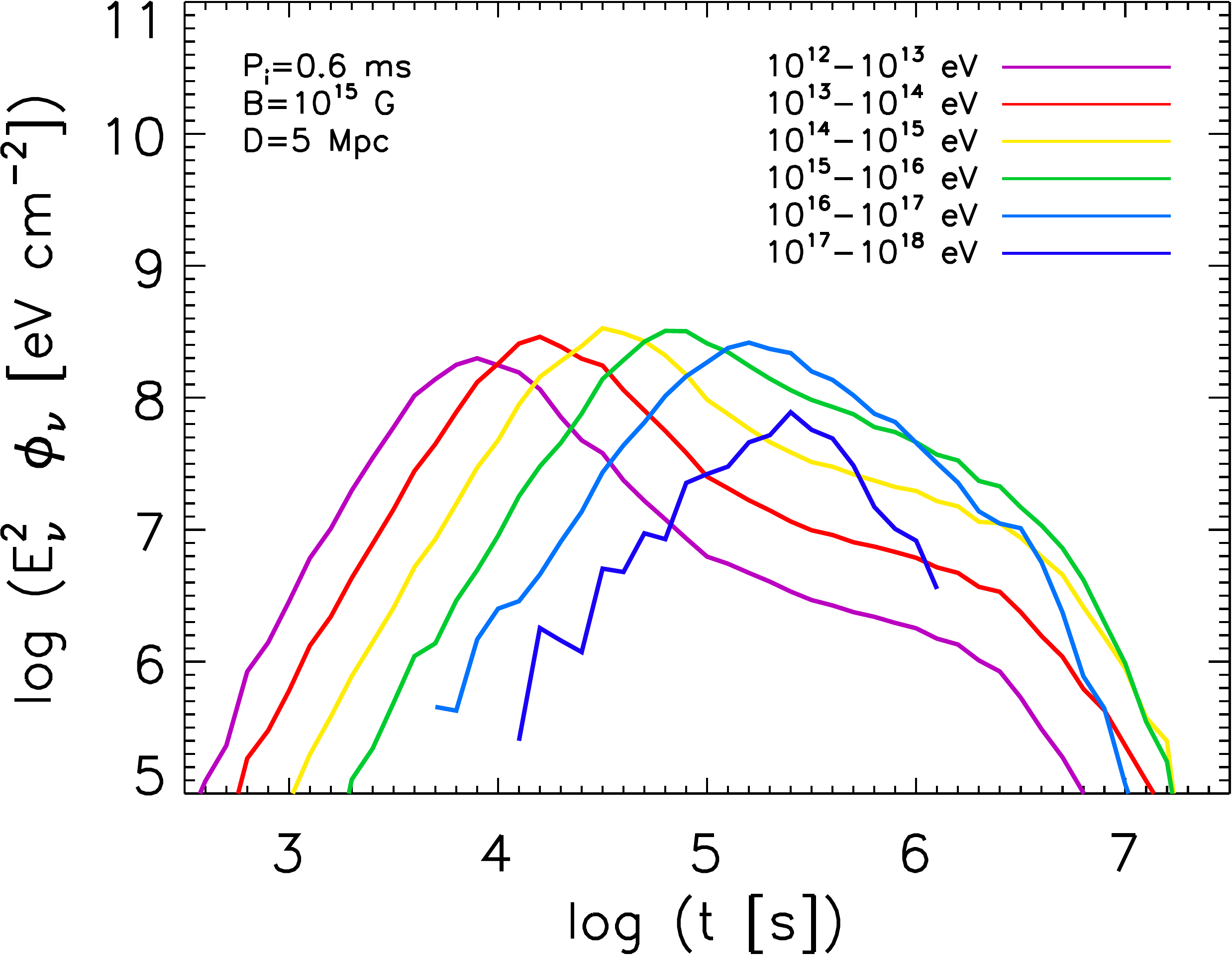,width=0.6\textwidth,clip=}
\epsfig{file=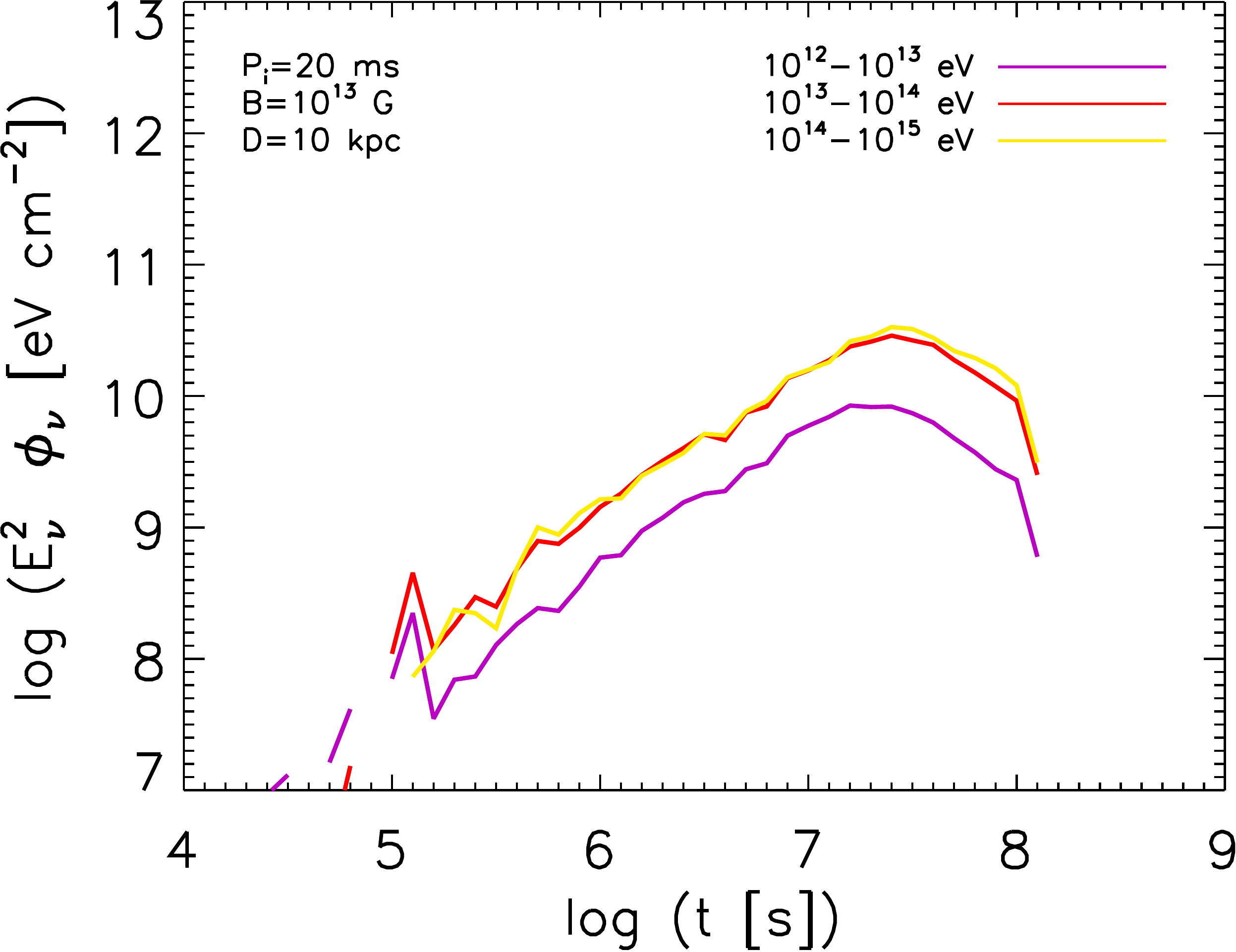,width=0.6\textwidth,clip=}
\caption{\label{fig:Ebin} Neutrino light curves of  the magnetar  case (top) and the  crab case (bottom). The input parameters for the two cases are listed in Table~\ref{table:input}. Pure proton injection has been assumed. Note that the color codes are different from Fig.~\ref{fig:spec}, and are  indicating different neutrino energies as listed in the legend box.
}
\end{figure}

The spin-down time of a pulsar ranges from minutes to thousands of years depending on the star's initial properties. The duration of  neutrino emissions could  last for the entire spin-down time in principle, but the majority of the neutrinos are produced before the environment gets too thin for interactions. 
In this section we assume that the observation could last for tens of years to study the neutrino spectrum accumulated over time. We   divide the detection period into logarithmic time bins then study the evolution of the spectrum  within each interval. 

The top panel in Fig~\ref{fig:spec} shows the expected neutrino spectrum from a newborn magnetar located at $5\,\rm Mpc$ with  $P_i=0.6\,\rm ms$ and $B =  10^{15}\,\rm G$ with pure proton injections. This is a highly magnetized and extremely fast-rotating neutron star that has a rare  birth rate. In particular, we note that  $P_i=0.6\,\rm ms$ roughly reaches  the minimum spin period allowed for neutron stars  \citep{1999A&A...344..151H, FKO12}. 
However, the short spin-down time $\tau_{\rm EM}\approx1000\,\rm s$ of the magnetar makes it a good representative of pulsars that release energy quickly and force cosmic rays to undergo severe interaction processes. 
We divide the observation time  from $10^{4}\,\rm s$ to $10^{6}\,\rm s$ into 5 logarithmic bins, indicated by corresponding colors as in the legend. In the first two bins centered at $10^4$ s and $10^{4.5}$ s, the ejecta is very dense with $\tau_{\pi p} \gg 1$, so the neutrino production is heavily suppressed,  as  pion products continue to suffer higher order $\pi p$ interactions.  The suppression continues till $ t_{\pi p} \approx\gamma_\pi\,\tau_\pi$, corresponding to a  peak  at $E_{\nu,\rm sup}(t) = 1.6\times10^{13}\,t_4^3\,\beta_{-1}^3\,M_{\rm ej,1}^{-1}\,\rm eV$.  Later at $10^5$ s and $10^{5.5}$ s, cosmic rays are accelerated to lower energy, so neutrinos peaked at $\min(E_{\nu}(t), E_{\nu,\rm sup}(t))$. 
When coming to the last bin centered at $10^6\,\rm s \sim \tau_{\rm thin}$,   the flux suffers from both decrease factors where the flux was truncated by $f_\pi<1$ and $E_\nu$ drops as a result of the spin-down. 
Unlike the injection spectrum of $E^{-1}$, the overall neutrino spectrum follows $\sim E^{-2}$,  because 
in such a fast-spinning magnetar, most cosmic rays were injected when the environment was still opaque, $\tau_{\rm EM} \ll \tau_{\rm thin}$. Thus, the spectrum was heavily boosted by secondary products. 
Intuitively, if a pion that  could have decayed into one primary neutrino with $E_{\nu,\rm prim}$ eventually fully interacted to $\sim E_{\nu,\rm prim} / E_{\nu,\rm sec}$ secondary neutrinos, each with energy $E_{\nu,\rm sec}$, the secondary flux would satisfy $(E_{\nu}\,dN/dE)_{\rm sec} = (E_{\nu}\,dN/dE)_{\rm prim} \times (E_{\nu,\rm prim} / E_{\nu,\rm sec})\propto E_{\nu,\rm sec}^{-1}$. Note that our result is quite different from Fig. 1 of \cite{Murase09} below $\sim 10\,\rm PeV$, as the neutrino spectrum  significantly changes when the secondary Np and $\pi p$ interactions are taken into account.

The spectrum and flux of neutrinos of hadronic origin mildly depend on the injection composition. In the middle panel of Fig.~\ref{fig:spec}, we present the neutrino emission from the magnetar case ($P_i=0.6\,\rm ms$, $B=10^{15}\,\rm G$), but with an injection composed of  pure iron (Fe) nuclei. Comparing it with the top panel one can tell that these two cases resulted in similar emission levels and spectrum shapes. Indeed, in a hardonuclear interaction with the center of momentum energy as high as  $\sim$TeV or above,  an Fe nucleus with energy $E_{\rm Fe}=56\,E_{\rm P}$ (see Eqn.~\ref{eqn:Emax}) is roughly equivalent to 56 protons each with energy $56/56\,E_{\rm P}$ when producing pions. On the other hand, as $\dot{N}_{\rm GJ} \propto 1/Z$, the injected numbers of Fe nuclei would be  26 times less than that of the protons.  As a consequence, an injection of nucleus with a mass number $A$ and charge $Z$ would result in $\sim A/Z\sim 2$ times of neutrino flux, but $\sim A/A=1$ times of $E_{\nu,\rm max}$, compared to a pure P injection. 

In the bottom panel of Fig.~\ref{fig:spec}, we show the neutrino spectrum from a Crab-like pulsar with  $ B=10^{13}\,\rm G$ and $P_i=20\,\rm ms$ at $10\,\rm kpc$. In this Crab case $\tau_{\rm EM}=1.2\times10^{10} B_{13}^{-2}P_{-1.7}^2\,\rm s \gg \tau_{\rm thin}$,  neutrinos are  produced by a very small fraction of cosmic rays that were accelerated before the ejecta became optically thin. We thus adjusted the observation time to $10^{6} - 10^{8}$ s. For $t\ll\tau_{\rm EM}$, $E_{\rm CR}=E(0)= 4.2\times10^{15}\,A\kappa_4^{-1}\eta_{-0.5}B_{13}P_{-1.7}^{-2}$ eV  and the production rate of particles   $\dot{N}_{\rm GJ}$ is a constant.  Thus, the neutrino emission in each time bin behaves similarly but with flux scaled to the size of the time window (as shown in the first three bins of Fig.~\ref{fig:spec} bottom).  When time increases to $t>\tau_{\rm thin}$ (the last two bins of Fig.~\ref{fig:spec} bottom), this feature  dominates over  the decreasing $f_\pi$. Compared to the previous case, pions and nuclei engaged in fewer interactions. With less boost from secondaries, the overall neutrino spectrum  roughly follows $E^{-1.5}$.

Figure~\ref{fig:Ebin} present the light curves of neutrinos from the newborn pulsars. Note now  the color codes have been changed to represent different neutrino energies. In the magnetar case (the top panel),  neutrinos at different energies display similar shapes in light curves, though those with higher energies show up later in time. For each energy interval, the light curve peaks at the time when $f_{\rm sup}=1$, specifically, $t_{\rm peak, lc} = 10^4\,E_{\nu, 13}^{1/3}\,P_{-3.2} \,\rm s$. This is consistent with the simulation result, where 10 TeV neutrinos peak around $10^4$ s and the peak time for neutrinos with higher energies increases by $E_\nu^{1/3}$. The decrement in the light curves at later times reflects  the decrease of the average collision number a particle would encounter during its escape. In particular, neutrinos above $10^{17}\,\rm eV$ are produced after the environment gets optically thin, thus the flux is suppressed by $f_\pi < 1$. In the crab case (the bottom panel), no neutrinos are produced above PeV due to the lack of cosmic ray acceleration. TeV neutrinos have less flux compared to those between 10 TeV and PeV, being consistent with Fig.~\ref{fig:spec}.

\section{Time-Dependent Neutrino Emissions} \label{Sec:flux}
If a nearby newborn pulsar can be resolved, its prompt neutrino emission would present a profile evolving with time.  We study  this emission profile in Section~\ref{Sec:flux_num}, and apply the scenario to young pulsars from recent supernova bursts in the local universe in Section~\ref{Sec:flux_past}.

\subsection{Flux level and Detectability}\label{Sec:flux_num}

Recall that at time t after a pulsar birth at distance D, neutrinos would emit at an intensity of 
\begin{equation}
E_\nu^2\Phi_\nu (t) = \frac{3}{8}\,E_{\rm CR}^2\,\frac{dN_{\rm CR}}{dEdt}\, \frac{1}{4\pi D^2}\,f_{\pi}(t)\,f_{\rm sup}(t)
\end{equation}
Assume that we are looking at the source at the age of $t_{\rm obs} = 1$ year,  below we estimate this intensity again  in the two cases of magnetar and crab. In the magnetar case, $t_{\rm ob}\gg\tau_{\rm EM}$, the typical neutrino energy reads $E_{\nu}=8.5\times10^{14}\,B_{15}^{-1}\,\eta_{-0.5}\,A\,\kappa_4^{-1}t_{\rm yr}^{-1}\,\rm eV$ and the flux would be
\begin{eqnarray}
E_\nu^2\,\Phi_\nu&=&2.8\times10^{-9}\,\left(\frac{B}{10^{15}\,\rm G}\right)^{-2}\, \left(\frac{P}{0.6\,\rm ms}\right)^2\, \left(\frac{t_{\rm obs}}{1\,\rm yr}\right)^{-4}\,
\left(\frac{D}{10\,\rm kpc}\right)^{-2}\\ \nonumber
&&\times\eta_{-0.5}\,A\,\kappa_4^{-1}\,Z^{-1} \,M_{\rm ej,1}\,\rm TeV\,cm^{-2}\,s^{-1}
\end{eqnarray}
In the crab case,  $t_{\rm ob}\ll \tau_{\rm EM}$, so that $E_{\nu} (t_{\rm obs})\approx E_{\nu}(0)=2.1\times10^{14}\,A\kappa_4^{-1}\eta_{-0.5}\,B_{13}\,P_{-1.7}^{-2}\,\rm eV$. The flux can be calculated as 
\begin{eqnarray}
E_\nu^2\,\Phi_\nu&=&4.4\times10^{-9}\,\left(\frac{P_i}{20\,\rm ms}\right)^{-4}\,\left(\frac{B}{10^{13}\,\rm G}\right)^2\,\left(\frac{t_{\rm obs}}{1\,\rm yr}\right)^{-2}\,
\left(\frac{D}{10\,\rm kpc}\right)^{-2}\\ \nonumber
&& \times \,\eta_{-0.5}\,A\,\kappa_4^{-1}\,Z^{-1} \,M_{\rm ej,1}\,\rm TeV\,cm^{-2}\,s^{-1}
\end{eqnarray}
In the crab case, the pulsar would take a long time to spin down, therefore at  a few years the neutrino flux only experiences the dilution of ejecta that  decreases with $t^{-2}$; while in magnetar case the neutrino flux would be further impacted by the pulsar spin-down, and  decreases with a faster rate of $t^{-4}$. 

\begin{figure}[h]
\centering
\epsfig{file=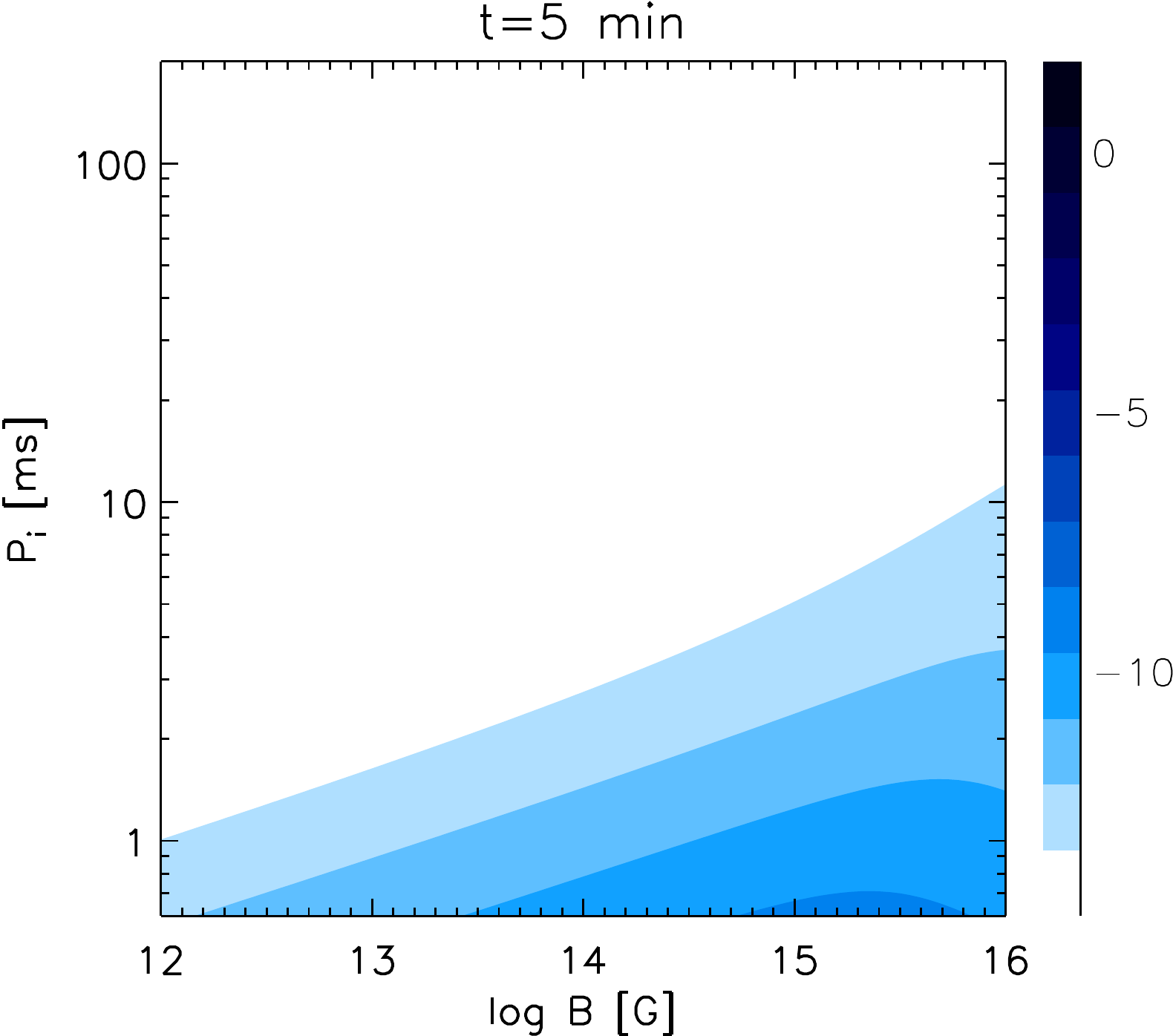,width=0.45\textwidth,clip=}
\epsfig{file=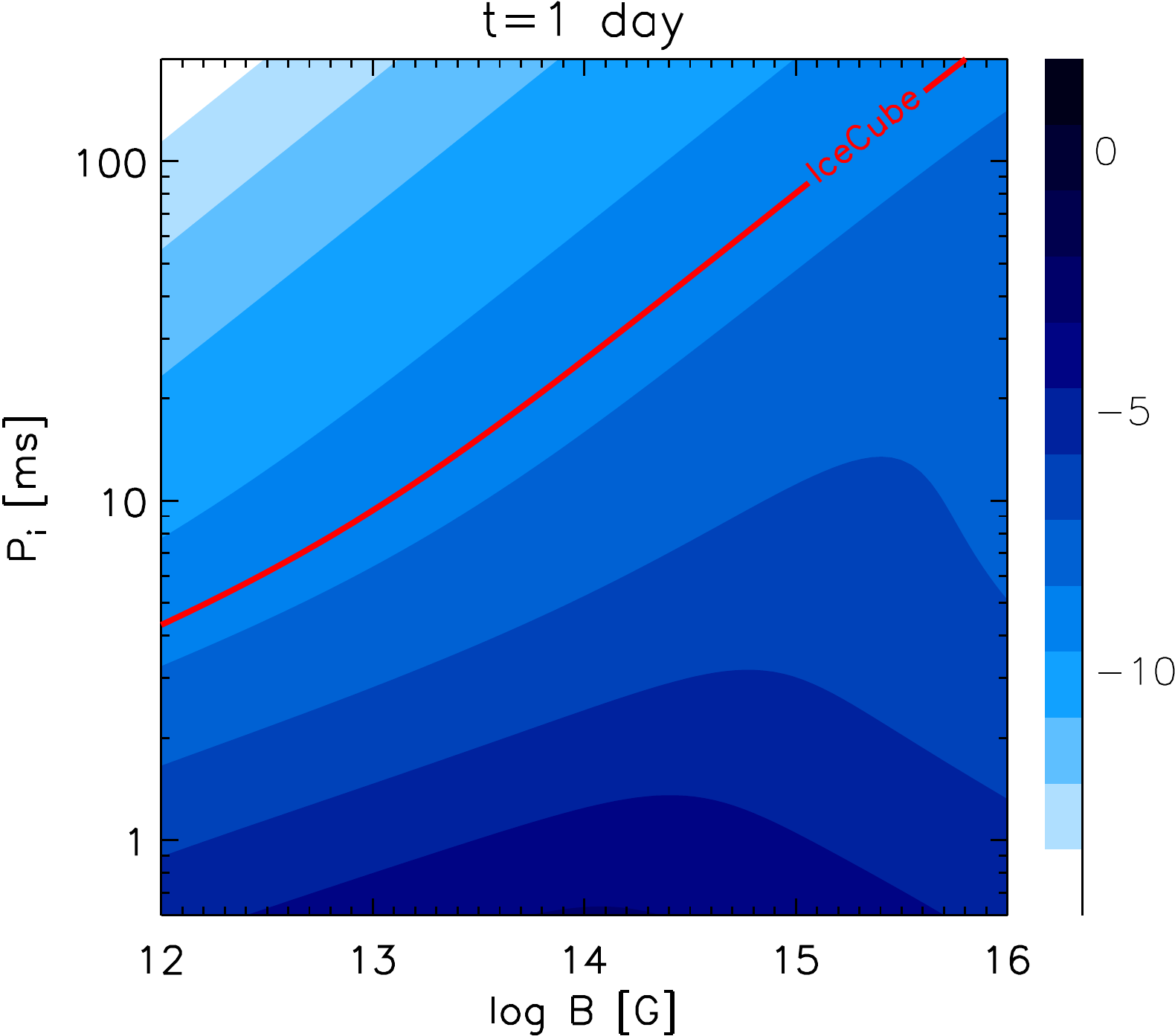,width=0.45\textwidth,clip=}
\epsfig{file=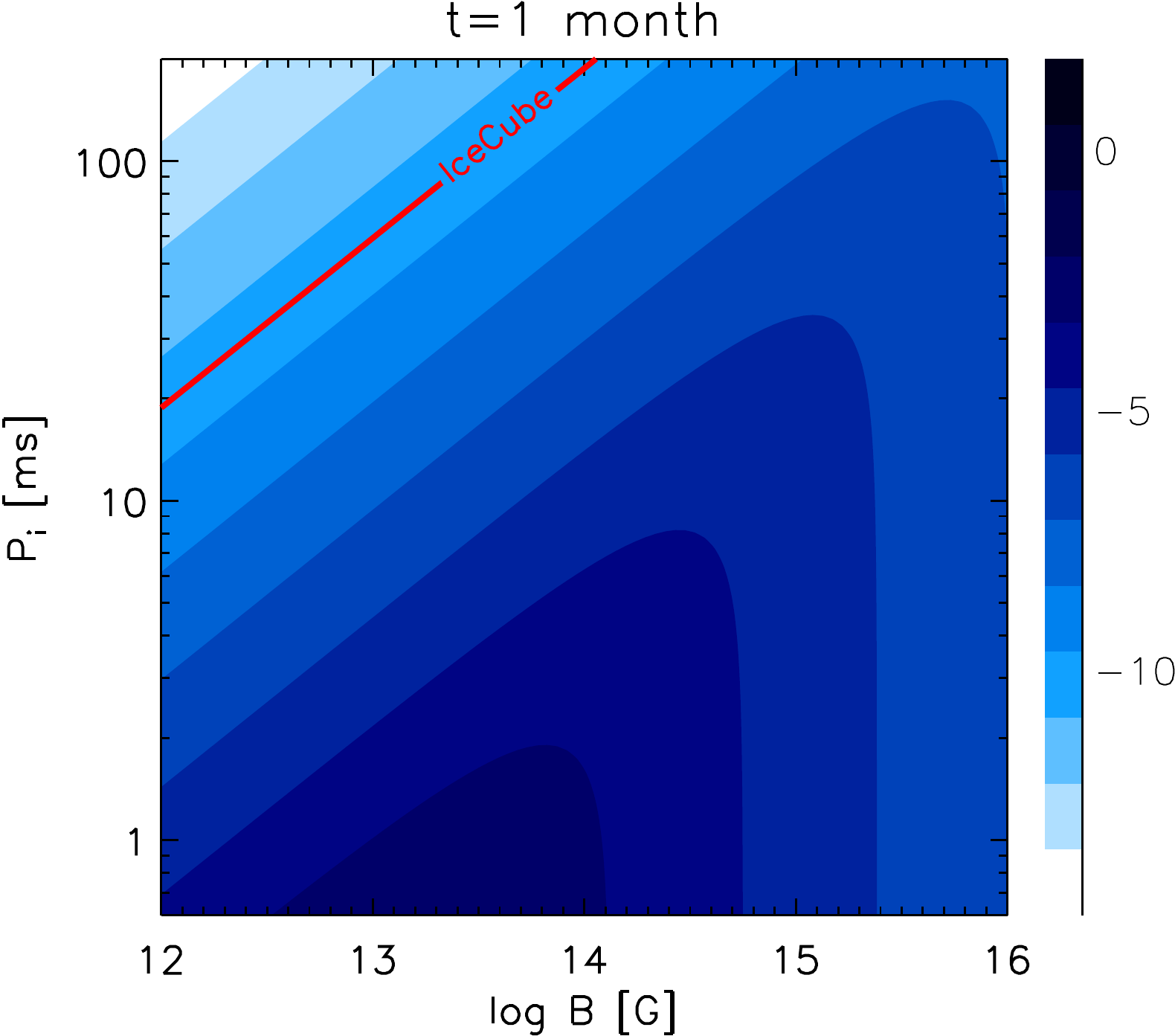,width=0.45\textwidth,clip=}
\epsfig{file=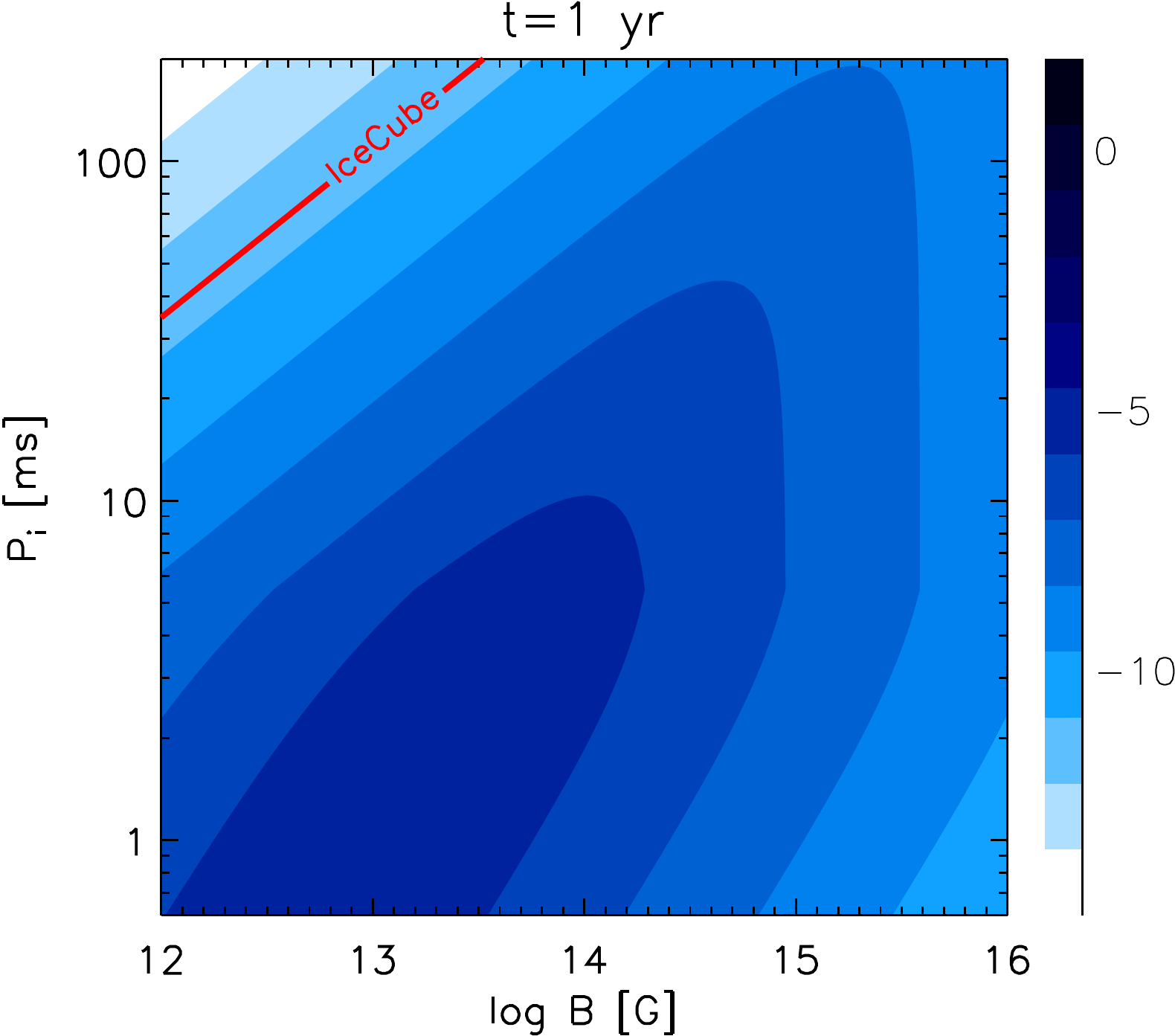,width=0.45\textwidth,clip=}
\epsfig{file=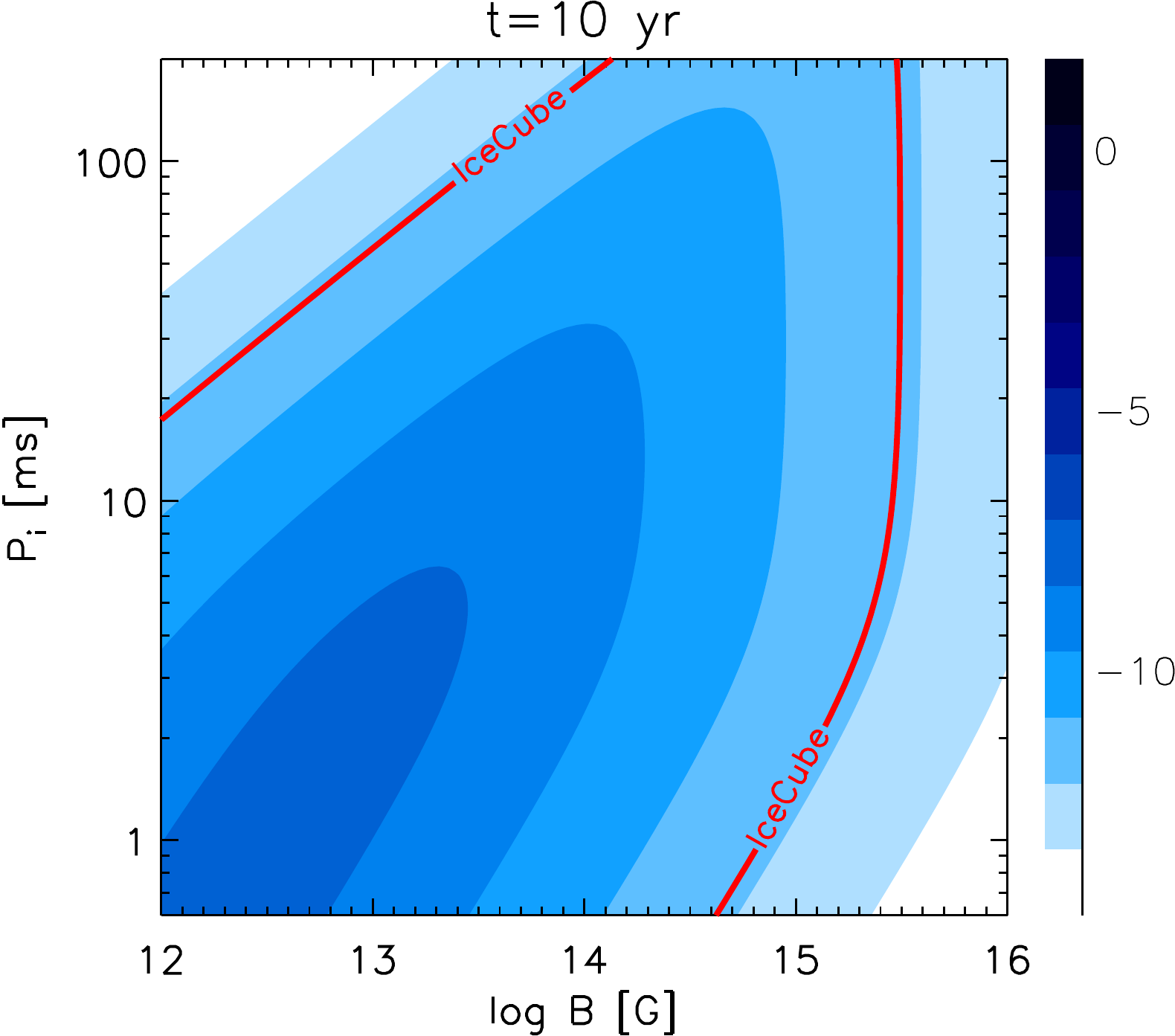,width=0.45\textwidth,clip=}
\epsfig{file=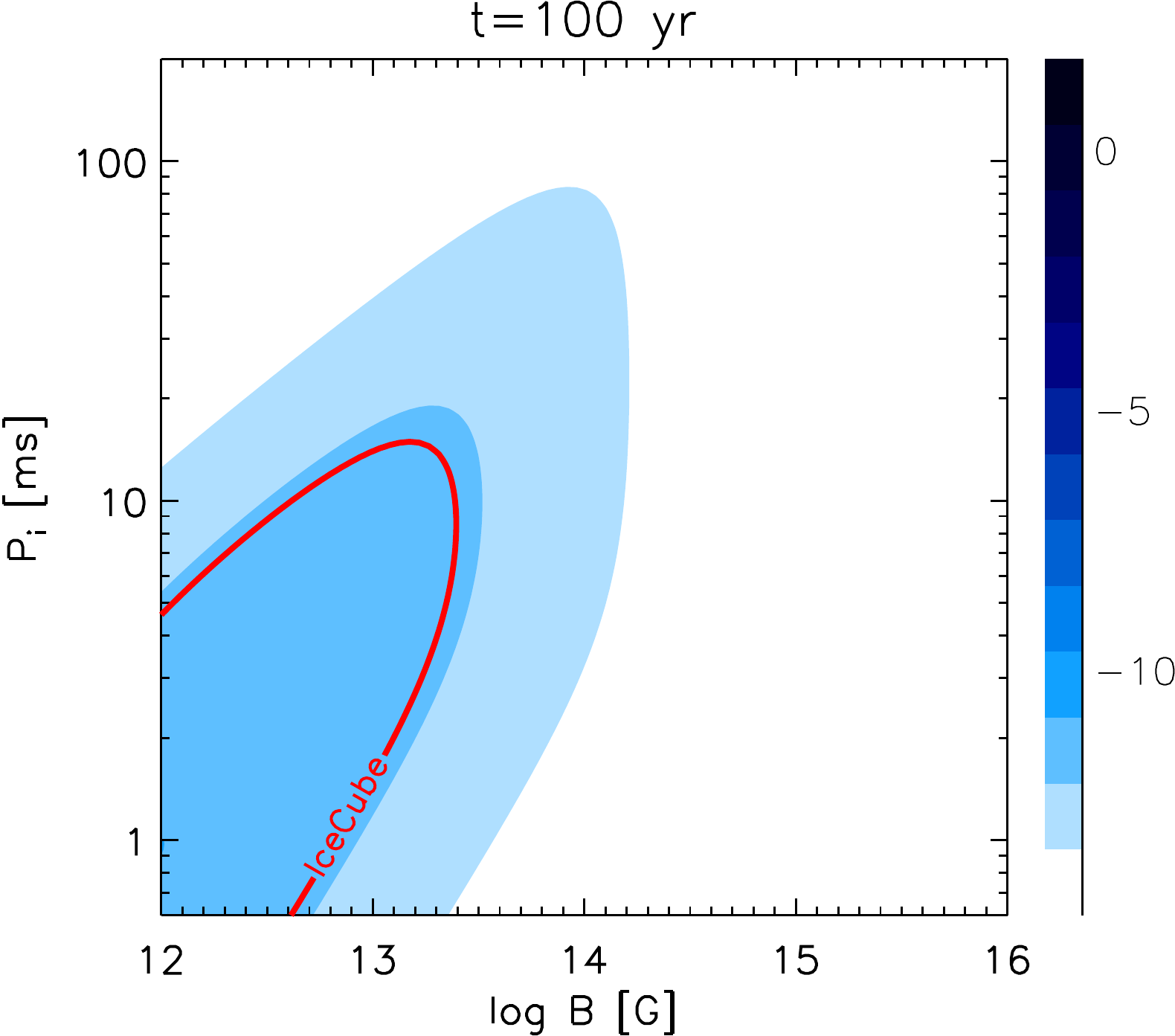,width=0.45\textwidth,clip=}
\caption{\label{fig:contour} Neutrino flux $E_\nu^2\Phi_\nu$ at t = 5 minutes, 1 day, 1 month, 1, 10 and 100 years after the  birth of a pulsar with $P_i, B$ scanning over the parameter space $(0.6\,\rm ms \le P_i \le 200\,\rm ms) \times (10^{12}\,\rm G\,\le  B \,\le 10^{16}\,\rm G)$. The flux is  in unites of $\rm TeV\,s^{-1}\,cm^{-2}$, shown in logarithmic scale with amplitude indicated in the color bar. The source distance is fixed at 10 kpc. Wind acceleration efficiency is set to be $\eta=0.3$.    The red solid line indicates the  medium point-source sensitivity of IceCube, $(E^2\,\Phi)_{\rm IC}= 10^{-12}\,\rm TeV\,s^{-1}\,cm^{-2}$ based on 4-year data \citep{Aartsen:2014cva} in the last two cases (at 10 and 100 years), but scaled to an observation time of  5 minutes, 1 day, 1 month, and 1 year in the top four cases. 
}
\end{figure}

To demonstrate the flux evolution over time, we present in Figure~\ref{fig:contour}  the neutrino flux level at 5 minutes, 1 day, 1 month, 1, 10 and 100 years after the birth of a pulsar, with $P_i, B$ scanning over the parameter space $(0.6\,\rm ms \le P_i \le 200\,\rm ms) \times (10^{12}\,\rm G\,\le  B \,\le 10^{16}\,\rm G)$. We fixed the source distance  at $D=10\,\rm kpc$.  The red line in the last two cases (at 10 and 100 years) represents IceCube's median sensitivity at $90\%$ C.L., $\sim 10^{-12}\,\rm TeV^{-1}\,cm^{-2}\,s^{-1}$ for energies between TeV - PeV in the northern sky based on 4 years of data \cite{Aartsen:2014cva}; this sensitivity is scaled to the corresponding observation time in the top four cases. At the very beginning, suppression from $\pi p$ interaction in the early environment is too fatal to allow the production of neutrinos. This suppression still distorts the flux at a few days and finally becomes less important after around 1 month (see equation~\ref{eqn:tpeak}). At about 1 month, the flux  reaches the maximum.  As the ejecta was optically thick with $f_\pi = 1$,  the flux depended only on the neutrino energy, scaling as $E_\nu^2 \propto \mu P_i^{-2}$. When $t \ge1 \,\rm yr$ the effect of the decrease of $f_{\pi}$ added in. Recall that for pulsars with high angular velocity that satisfies $1/2\,I\,\Omega_i^2 \gg 10^{51}\,\rm erg$, ejecta expands with speed  $\beta \propto \Omega_i$. Thus in the bottom half of the $t=1\,\rm yr$ plot, $E_\nu^2\Phi_\nu \propto E_\nu^2\,f_{\rm pp} \propto \mu^{-1}\,P_i^2$. This shape maintained to $t=10\,\rm yr$. Finally at $t=100\,\rm yr$, the interaction probability decreased to $f_\pi\sim 10^{-4}$. In addition, due to the spin down the majority of pulsars    failed to produce neutrinos above 1 TeV, to reach the optimized observation window of the IceCube Observatory \citep{Achterberg:2006md}.  

\subsection{Existing nearby pulsars}\label{Sec:flux_past}

\begin{table}[t]
\caption{Properties of Pulsars} \label{table:pulsar}
\centering
\begin{tabular}{ccccccccc}
\hline\hline 
PSR \T& SNR & Supernova & D & $P_i$ & B  & Age & $E_{\rm exp}$ & $M_{\rm ej}$\\ 
 &   & Type & (kpc) & (ms) & ($10^{12}$ G) & (yr) &  ($10^{51}\,\rm erg$) & ($M_\odot$)\B  \\
\hline
B0531+21 \T&Crab & IIP & 2 & 20 & 4 & 950 & 1 & 9.5 \\
J0205+64 & 3C 58 & IIP & 3.2 & 50 & 4 & 2400 & 1 & 3.2 \\  
J1846-03 & Kes 75 &Ib/c & 19 & 30 & 48 & 1000 & 2.1 & 16.4 \\
B1509-58 & MSH 15-52 &Ib/c & 5.2 & 10 & 14 & 1700 & 7 & 4 \B\\ 
\hline
\end{tabular}
\end{table}

\begin{figure}[h]
\centering
\epsfig{file=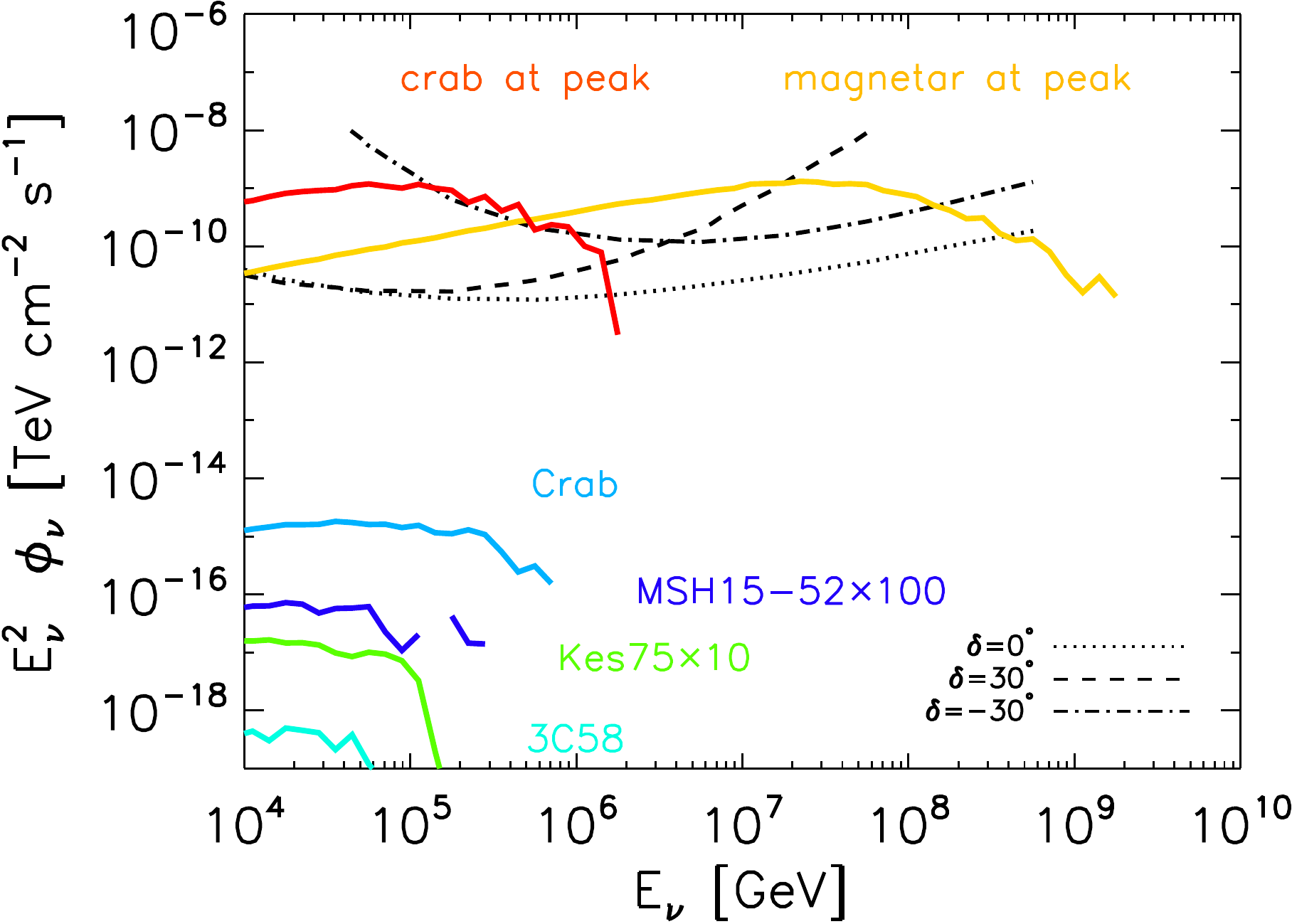,width=0.8\textwidth,clip=}
\caption{\label{fig:event} Expected neutrino spectra of four nearby young pulsars at their current ages, compared to the $5\sigma$ differential discovery  sensitivities of  the IceCube Observatory \cite{Aartsen:2014cva}. As indicated in the legends, the pulsars include   Crab pulsar, J0205, J1846 and B1509, with their properties summarized in Table~\ref{table:pulsar}. Note that the flux of MSH15-52 and Kes 75 was enlarged by 100 and 10 times accordingly to not overlap with other lines. Also plotted are the predicted emission from a fast-spinning magnetar  located at 5 Mpc at its neutrino peak time $t_{\rm peak} = 2.4\times10^5\,\rm s$ and a Crab-like pulsar at $10\,\rm kpc$ at $t_{\rm peak}=2.1\times10^5\,\rm s$. Their input parameters are listed in Table~\ref{table:input}. }
\end{figure}

Four nearby young pulsar wind nebula and supernova remnants in which central pulsars have been identified are listed in Table~\ref{table:pulsar}. This list includes all the objects within 5 kpc and with estimated initial spin period  less than $50\,\rm ms$ in Table 1 of  \cite{Chevalier05} except for Kes 75. Kes 75 is shown because of its large magnetic field.  These objects are known to actively interacting with ejecta  \cite{Chevalier05, 2010arXiv1005.1831B}. 

In Figure~\ref{fig:event} we show the expected neutrino emission from these pulsars at their current ages. Among the list, neutrino flux from Crab pulsar has the highest flux, due to its young age and close distance.  3C58 has the lowest emission, as a result of the old age and the relatively slow spinning speed. For comparison, we also show the  $5\sigma$ differential discovery potentials of IceCube at declination $0^\circ, \, -30^\circ\,\rm and \,+30^\circ$ \citep{Aartsen:2014cva}. The expected neutrino emissions from nearby young pulsars are far below the discovery sensitivities, consistent with the non-detection of point sources \citep{Aartsen:2014cva}. As a comparison, we show the neutrino flux from two future pulsars  at their peak time, with yellow and red lines indicating magnetar  and crab case respectively. These two cases would be  easily observed by IceCube. 

\section{Birth rate of detectable pulsars}\label{Sec:flux_rate}
In this section we investigate the birth rate of pulsars that would be detectable by the IceCube Observatory.  
Assuming that a pulsar could at least be observed at the peak time $t_{\rm peak}$ as defined in equation~\ref{eqn:tpeak}, we define  the maximum distance of a detectable pulsar by
\begin{equation}
D_{\rm max} =( (E_\nu^2\,dN/dEdt)_{\rm peak}/4\pi/(E_\nu^2\Phi_\nu)^{\rm det})^{1/2} 
\end{equation}
with $(E_\nu^2\Phi_\nu)^{\rm det}$ the   detector sensitivity. 

The median IceCube sensitivity reads $ (\Phi_\nu)^{\rm IC} =10^{-11}\,\rm TeV^{-1}\,cm^{-2}\,s^{-1}$ for energies above 100 TeV and  $ (\Phi_\nu)^{\rm IC} =10^{-12}\,\rm TeV^{-1}\,cm^{-2}\,s^{-1}$ for energies between 1 TeV - 1 PeV \citep{Aartsen:2014cva}. Taking this sensitivity, $D_{\rm max}$ can be estimated to be, in the magnetar case,
\begin{eqnarray}\label{eqn:Dmax_mag}
D_{\rm max}=87.2\,B_{15}^{-3/4}\,\eta_{-0.5}^{1/4}\,A^{1/4}\,\kappa_4^{-1/4}\,Z^{-1/2}\,M_{\rm ej,1}^{-1/4}\,P_{i,-3.2}^{-3/4}\left(\frac{\Phi_\nu^{\rm det}}{\Phi_\nu^{\rm IC}}\right)^{1/2}\,\rm Mpc
\end{eqnarray}
and in the crab case,
\begin{eqnarray}\label{eqn:Dmax_crab}
D_{\rm max}=0.5\,B_{13}\,P_{-1.7}^{-2}\,\eta_{-0.5}^{1/2}\,A^{1/2}\,\kappa_4^{-1/2}\,Z^{-1/2}\left(\frac{\Phi_\nu^{\rm det}}{\Phi_\nu^{\rm IC}}\right)^{1/2}\,\rm Mpc
\end{eqnarray}

According to \cite{Faucher06}, the initial spin periods and magnetic fields of newborn isolated neutron stars displace  a distribution $f(P_i, B)$,  with   $P_i$  normally distributed, centered at $\langle P_i \rangle = 300\,\rm ms$ with  standard derivation  $\sigma_{P_i}=150\,\rm ms$;  B  log-normally distributed with $\langle\log B\rangle=12.65$ and $\sigma_{\log B} = 0.55$. A lower limit of the initial spin period, $P_{i,\rm min}\sim 0.6-1\,\rm ms$ is required for  a stable configuration of a neutron star \citep{1999A&A...344..151H,2006Sci...311.1901H}. In this work, we set $P_{i,\rm min}$ to be $0.6\,\rm ms$.

 Based on the pulsar  distribution,  below we investigate the birth rate of pulsars that could be detected by the IceCube Observatory. 
We first investigate the detection upper limit in Sec.~\ref{subsec:upperlimit}, and then refine this upper limit by considering explicitly  the contribution from pulsars in nearby clusters  (Sec.~\ref{subsec:cluster}) and pulsars in the Local Group (Sec.~\ref{subsec:localGroup}).

\subsection{Upper limit}\label{subsec:upperlimit}

The local core-collapse supernova has a formation rate of $\rho_0 = 10^{-4}\,\rm Mpc^{-3}\,yr^{-1}$ \citep{Izzard:2003uz}. Taken this rate, we can calculate the birth rate of the detectable pulsars in the universe by counting the formation rate of each pulsar in its own maximum volume and summing the population over the (P, B) probability space:
\begin{eqnarray}\label{eqn:extragalRate}
\bar{\Re}_{\rm limit}&=&\int dP_i\, \int dB\, \rho_0\frac{4\pi}{3}D_{\rm max}(P_i, B)^3 w(P_i, B) f(P_i, B)  
\end{eqnarray}
Here we only count the pulsars that could contribute to the IceCube sensitivity window by setting a window function $w(P_i, B)= \mathbbm{1}(E_\nu\ge \rm TeV)$.

\begin{figure}[h]
\centering
\epsfig{file=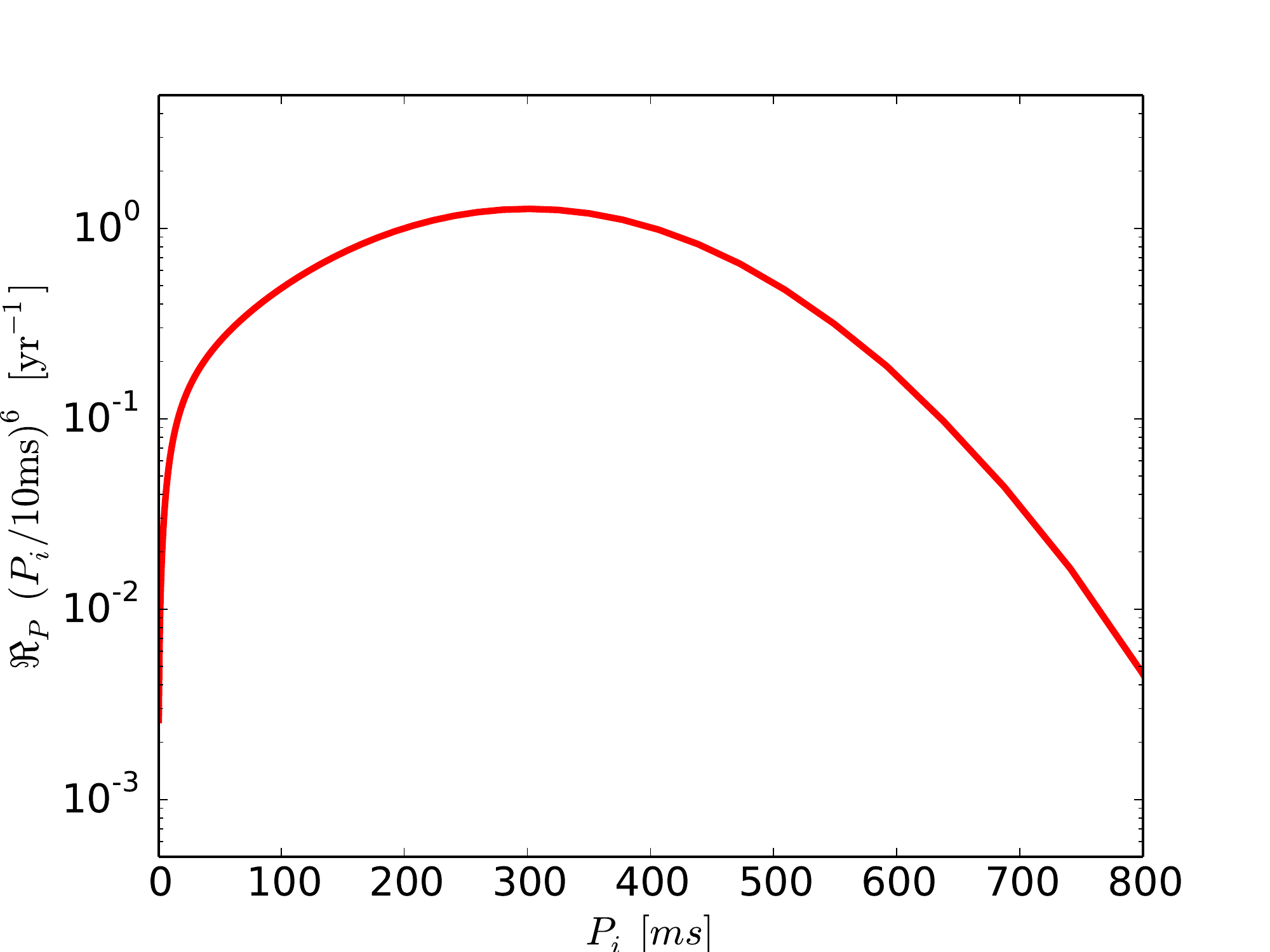,width=0.8\textwidth,clip=}
\caption{\label{fig:extragalRate} Dependence of the upper limit of birth rate of detectable pulsars  on the initial spin periods $P_i$.  $\Re_P$ is defined in Eqn.~\ref{eqn:Re_Pi} which has marginalized  over the magnetic field distribution. Note that the y-axis is scaled by $\left(P_i/10\,\rm ms\right)^6$ for presentation purposes (see Sec.~\ref{subsec:upperlimit} for more details).  }
\end{figure}

$\bar{\Re}_{\rm limit}$ significantly depends on the minimum spin periods $P_{i,\rm min}$ and the acceleration efficiency $\eta$.  To understand its dependence on $P_{i,
\rm min}$, in Fig.~\ref{fig:extragalRate} we show $\Re_P$ as a function of $P_i$, with $\Re_P$ denoting the inner integral of Eqn.~\ref{eqn:extragalRate}  which marginalizes over the magnetic field distribution:
\begin{eqnarray}\label{eqn:Re_Pi}
\Re_P(P_i) = \int dB\, \rho_0\frac{4\pi}{3}D_{\rm max}(P_i, B)^3 w(P_i, B) f(P_i, B) 
\end{eqnarray}
With the y-axis of Fig.~\ref{fig:extragalRate} scaled  by $(P_i/10\,\rm ms)^6$, one could easily find that $\Re_P P_i^6$ roughly agrees with the $P_i$ distribution, which is a Gaussian distribution with a mean of  300 ms and a standard derivation of 150 ms.  The reason is that as the magnetic field distribution strongly peaks at $10^{12-13}\,\rm G$, $\Re_P$ is mostly contributed by Crab-like pulsars which have $D_{\rm max} \propto P^{-2}$, thus $\Re_P\propto D_{\rm max}^3\,f(P_i)\propto f(P_i)\,P_i^{-6}$. $\Re_P\,P_i^6$ cuts off sharply at low $P_i$ further   due to the contribution of fast-spinning magnetars,  which have a particularly large $D_{\max}$ but scaling to $P_i^{-3/4}$. 
Fig.~\ref{fig:extragalRate} indicates that the birth rate of observable extragalactic  pulsars is heavily dominated by the fast-spinning pulsars, and thus strongly depends on $P_{i, \rm min}$. Moreover, the upper limit depends on the wind acceleration efficiency  by $\bar{\Re}_{\rm limit}\sim \eta^{1.3}$
due to the weighted contribution from magnetars and crabs.  Results of $\bar{\Re}_{\rm limit}$ with different inputs of $P_{i,\rm min}$ and $\eta$ are listed in Table~\ref{table:extragalRate}.  Considering $P_{i,\rm min} = 0.6\,\rm ms$ and $\eta=0.3$,  we find  $\bar{\Re}_{\rm limit}= 0.29\,\rm yr^{-1}$.

\begin{table}[h]
\caption{Upper Limits on the Birth Rate of Detectable Pulsars in $\rm year^{-1}$} \label{table:extragalRate}
\centering
\begin{tabular}{c|ccc}
\hline\hline 
\backslashbox{$P_{i,\rm min}$}{$\eta$} & 0.1 & 0.3 & 0.7 \\ 
\hline
0.6 ms & 0.07 & 0.29 & 0.82 \T \\
1 ms & 0.01 & 0.04 & 0.12 \B \\ 
\hline
\end{tabular}
\end{table}

Notice that $\bar{\Re}_{\rm limit}$ is an optimized upper limit because: i) although $t_{\rm peak}$ depends on $P_i$ and $B$  and varies from months to tens of years, pulsars have been assumed be always detectable at their brightest point; ii) \cite{FKO13} found a normalization factor $f_s\approx 5\%$ if using UHECR measurements to confine the pulsar population, which means $\sim 5\%$ of the pulsar population is required to have the right configuration to successfully inject and  accelerate ions. If taking this confinement, the upper limit would  drop to  $5\%$ of what calculated from Eqn.~\ref{eqn:extragalRate}. 
 iii) $\bar{\Re}_{\rm limit}$  has ideally assumed a  continuous pulsar distribution up to $\sim100$ Mpc in the local universe, while in reality the local large-scale structure is not uniform.  We will further demonstrate this point next. 

\subsection{Pulsars in nearby clusters} \label{subsec:cluster}
To take into account that the local large-scale structure is not uniform,  we refine our estimation of the extragalactic pulsar detection rate by considering explicitly the contribution from nearby clusters. 
For a nearby cluster at $D_{\rm cluster}$ with a total pulsar birth rate of $\rho_{\rm cluster}$, the rate of pulsars that can be observed from such a distance can be estimated by
\begin{eqnarray}
\bar{\Re}_{\rm cluster}=\rho_{\rm cluster}\,\int w(P_i, B) P_{\rm obs}^{\rm cluster}(P_i, B)\, f(P_i, B)dP_idB
\end{eqnarray}
$P_{\rm obs}^{\rm cluster}$ is the observational chance of a pulsar inside the cluster. Since a pulsar can be observable only when its maximum distance exceeds the distance of the cluster, we have
\begin{equation}
P_{\rm obs}^{\rm cluster} = \begin{cases} 1 &\mbox{if } D_{\rm max} \geq D_{\rm cluster} \\ 
0 & \mbox{if } D_{\rm max} < D_{\rm cluster}. \end{cases} 
\end{equation}

Take the closest Virgo Cluster at $D_{\rm Virgo}\sim 16.5$ Mpc for example. Within its total newborn pulsar population which has a rate of $\sim 20\,\rm yr^{-1}$  \citep{2005ApJ...634L.165S},  $0.16\%$  have $D_{\rm max} \geq D_{\rm Virgo}$ and thus can be seen from the Earth. This corresponds to a birth rate of $0.032 \,\rm yr^{-1}$. 
Note that we have assumed $P_{i,\rm min}=0.6\,\rm ms$, $\eta=0.3$, and that the sources are from the central region of the Virgo Cluster.

With the same calculation, we find that the other nearby clusters -- Centaurus Cluster at 52 Mpc, Perseus Cluster at 74 Mpc, and   Coma Cluster at 102 Mpc have $0.058\%$, $0.039\%$, and $0.025\%$ of their local pulsars observable. Along with the Virgo Cluster, and assuming each of them has a local core-collapse SN rate of $20\,\rm yr^{-1}$, we get  a total detection rate of 0.06 per year from pulsars in the nearby clusters. 

\subsection{Pulsars in the Local Group}\label{subsec:localGroup}
Observations suggest that a total of 4.5 core-collapse supernova are expected per century in the Local group, including  2.5 from our Galaxy ($\rho_{\rm Gal} = 0.025\,\rm yr^{-1}$), 0.5 from the Magellanic Clouds, and 1.5  from M31 plus M33 \citep{1994ApJS...92..487T, Faucher06, 2013ApJ...778..164A}.

We first consider the birth rate of observable Galactic newborn pulsars. A pulsar would definitely be observed if it has $D_{\rm max}$ greater than the size of our Galaxy, $R_{\rm Gal}\approx 15\,\rm kpc$. On the other hand, if a pulsar has  $D_{\rm max} < R_{\rm Gal}$, it still has a  probability of $(D_{\rm max}/R_{\rm Gal})^3$ to be observed, with a very rough assumption that  the occurrence of core-collapse supernova is  uniform inside the Galaxy. The above two cases summarize to an observation probability $P_{\rm obs}^{\rm Gal}(P_i, B) \equiv \max((D_{\rm max}/R_{\rm Gal})^3, 1)$, which yields a total birth rate of detectable pulsars in the Galaxy to be 
\begin{eqnarray}
\bar{\Re}_{\rm Gal}&=&\rho_{\rm Gal}\,\int w(P_i, B) P_{\rm obs}^{\rm Gal}(P_i, B)\, f(P_i, B)dP_idB\\ \nonumber
&=& 0.01\,\rm yr^{-1}
\end{eqnarray}
assuming that $P_{i,\rm min}=0.6\,\rm ms$ and $\eta=0.3$. 
The same calculation can be applied to the birth rate of pulsars in the Local Group, which has a size of  $R_{\rm LC} \approx 1\,\rm Mpc$ \citep{LocalGroup}. However, due to the larger distance, nearby galaxies only enhance $\bar{\Re}_{\rm Gal}$ by $4.8\times10^{-4}\,\rm yr^{-1}$.

\section{Conclusions and Discussion}\label{Sec:discussion}

In this work we have investigated high energy neutrino emissions from individual newborn pulsars  in the Local Universe. 
Through both theoretical and numerical methods, we   examined the hadronuclear interactions between   cosmic rays and the  source ambience left over from the supernova explosion and reported on the detectability and emission profiles of the high-energy neutrino products.   We   found that 

\vspace{1em}
I) unlike the $E^{-1}$ spectrum of the injection cosmic rays, the energy spectrum of the neutrinos  has a softer shape, following $\sim E^{-2}$ for a fast-spinning magnetar, and  $\sim E^{-1.5}$ for a Crab-like pulsar. The modifications to the spectrum mainly originate from the high-order interactions of secondary ions and charged pion products in the early environment; 

\vspace{1em}
II) The flux level and shape of time-integrated neutrino spectrum  are not significantly sensitive to the chemical composition of the injected cosmic rays;

\vspace{1em}
III) Applying the scenario to known nearby pulsars found their neutrino detectability to be below the sensitivities of the current generation of detectors and consistent with the negative results of point  source searches;

\vspace{1em}
IV)   The birth rate of pulsars with neutrino emissions detectable by the IceCube Observatory is constrained with an upper limit of 0.29 per year. In particular, the birth rate of detectable Galactic pulsars is estimated to be 0.01 per year, and that of pulsars from nearby clusters is found to be 0.06 per year. 

\vspace{1em}
One caveat of our work is that we did not take into account the  radiation fields of the pulsar. A relativistic Fermi process is known to occur in the termination shock of the pulsar wind nebula \citep{1996MNRAS.278..525A, Kirk:2007tn}. Nonthermal radiations from electron and positions accelerated in the termination shocks are  expected to lead to photodisintegration of the cosmic rays,  in addition to the hadronuclear channel discussed in this work.  Figure~4 of  \cite{LKP13} compared the neutrino emission from $ p\gamma$ and $pp$ interactions to show
 that for a population of pulsars with identical parameters, the neutrino production from $p\gamma$ interactions is 2-3 orders less than that from $pp$ interactions, suggesting that the $p\gamma$ channel is sub-dominant to the $pp$ channel in neutrino productions. Moreover, in extremely radiative magnetars, a fraction of the outflow energy may dissipate as thermal radiation via the shocks \citep{Thompson:2004wi,Metzger:2006mw, Bucciantini:2006ra, 2010ApJ...717..245K}. The photomeson production rate with these thermal photons is $\tau_{p\gamma,\rm th} = 12\,E_{52}^{3/4}\,\eta_{\gamma,-1}^{3/4}\,t_{\rm yr}^{-5/4}\,\beta_{-1.5}^{-5/4}$, where $\eta_\gamma \sim0.1$ is the fraction of total pulsar energy that goes into thermal photons \citep{Murase09}.  Notice that $\tau_{p\gamma,\rm th}$ decreases slower than $\tau_{\rm Np}$ with  $\tau_{p\gamma,\rm th}\propto t^{-5/4}$,  so  the thermal photons could produce neutrinos  comparable to those with hadronic origins after about a year.  However, thermal photons are not as important in less radiative pulsars like Crab.

When considering the cosmic ray interaction with the ejecta, we ignored the delay of cosmic rays in the magnetic field of pulsar wind nebula. The larmor radius of particles reads $ r_L \approx 10^{11}\,(E/1\,\rm PeV)\,Z^{-1}\,(B_{\rm pwn}/30\, G)^{-1}\,\rm cm$, in a nebula with field strength $B_{\rm pwn}\sim37\,P_{-3}^{-5/2}\,B_{13}^3\,(t/\tau_{\rm EM})^{-7/4}\,\rm G$  \citep{LKP13}. At small t,  the size of the supernova ejecta is not significantly greater than $r_L$, then   particles escape recti-linearly with negligible time delay.  As time increases    $R_{\rm ej} \gg r_L$,  particles diffuse in the structure and exit with a deflection of order unity, $\delta \theta \approx (1+2\,r_L^2/R\lambda)^{-1} \sim 1^\circ$, and a time delay comparable to the crossing time, $\delta t = R\delta \theta^2/6c \sim R/c $ \citep{Kotera:2008ae}. This adds to the pion production rate $f_\pi$ a factor of $\sim 2$. However, this little increase in $f_\pi$ would be dominated  by the fast decrease of the number density of  target baryons in the ejecta. Therefore the effect from the magnetic field of the pulsar wind nebula is considerably small. 

One aim of the present work is to provide an observational template for neutrino emissions from future  pulsar sources in the local universe. Compared to other transient sources like Gamma-ray Bursts, fast-spinning pulsars display unique features  in high-energy neutrino emissions, including time-variable spectrum shapes, insensitivity to the chemical composition of injected parent particles, and an   $\sim E^{-1.5}$ spectrum index for the majority of the population. Such neutrinos might be detected by   current and future generations of high-energy neutrino observatories like IceCube \citep{Aartsen:2014cva}, KM3NeT \citep{KM3NeT}, and by future ultrahigh energy neutrino detectors like JEM-EUSO \citep{Adams:2012tt}, Askaryan Radio Array (ARA) \citep{2012APh....35..457A}, and the Antarctic Ross Ice-Shelf ANtenna Neutrino Array (ARIANNA) \citep{Barwick:2006tg}, thereby providing a unique way for understanding particle acceleration  in a newborn pulsar. 





\acknowledgments
The author would like to thank the anonymous referee for helpful comments and suggestions.
The author acknowledges S. Dodelson, K. Kotera, N.  Neilson, A.   Olinto, P. Privitera and S.   Wakely   for very fruitful discussions. This work was supported by the NSF grant PHY-1068696 at  the University of Chicago, and the Kavli Institute for Cosmological Physics through grant NSF PHY-1125897 and an endowment from the Kavli Foundation.
KF acknowledge financial support from NASA 11-APRA-0066 at the University of Chicago.
This work made use of computing resources and support provided by the Research Computing Center at the University of Chicago.

\bibliography{F14}

\end{document}